\documentclass[traditabstract]{aa}%
\usepackage{graphicx}
\usepackage{txfonts}
\usepackage{supertabular}
\usepackage{textcomp}
\usepackage{booktabs}
\usepackage{lscape}
\usepackage{color}
\usepackage{natbib}
\usepackage{xspace} 
\usepackage{tabularx}
\usepackage{pst-all}

\graphicspath{{./}}
\usepackage{txfonts}
\def\C{3C\,120\xspace}
\def\Ha{H$\alpha$\xspace}
\def\Hb{H$\beta$\xspace}
\def\V16{BMT\xspace}
\def\Vy6{RoBoTT\xspace}
\def\BII{BEST\,II\xspace}

\def\SCAMP{\it Scamp\rm\xspace}
\def\SWARP{\it Swarp\rm\xspace}

\def\USB{Universit\"atssternwarte Bochum\xspace}

\def\mdot{$\dot{M}$\xspace}
\def\rblr{$R_{\mathrm{BLR}}$\xspace}
\def\rdust{$R_{\mathrm{dust}}$\xspace}
\begin{document}
%
%

\title{Simultaneous H$\alpha$ and dust reverberation mapping of 3C\,120: Testing the bowl-shaped torus geometry}

\author{Michael Ramolla
  \inst{1}
  \and
  Martin Haas
  \inst{1}
  \and
  Christian Westhues
  \inst{1}
  \and
  Francisco Pozo Nu\~nez
  \inst{1}
  \and
  Catalina Sobrino Figaredo
  \inst{1}
  \and \\
  Julia Blex
  \inst{1}
  \and
  Matthias Zetzl
  \inst{2}
  \and
  Wolfram Kollatschny
  \inst{2}
  \and
  Klaus W. Hodapp
  \inst{3}
  \and
  Rolf Chini
  \inst{1,4}
  \and
  Miguel Murphy
  \inst{4}
}
\institute{Astronomisches Institut, Ruhr--Universit\"at Bochum,
  Universit\"atsstra{\ss}e 150, 44801 Bochum, Germany, {email: ramolla@astro.rub.de}
  \and
  Institut f\"ur Astrophysik, Universit\"at G\"ottingen,
  Friedrich-Hund Platz 1, 37077, G\"ottingen, Germany
  \and
  Institute for Astronomy, 640 North A'oh\={o}k\={u} Place, Hilo, HI 96720-2700, USA
  \and
  Instituto de Astronom\'{i}a, Universidad Cat\'{o}lica del Norte, Avenida Angamos 0610, Antofagasta, Chile
}

\date{Received 11 October 2017; accepted 8 August 2018}

\abstract{ We monitored the Seyfert-1 galaxy \C between September 2014
  and March 2015 at the \USB near Cerro Armazones in $BVRIJK$ and a
  narrowband filter covering the redshifted \Ha line.  In addition we
  obtained a single contemporary spectrum with the spectrograph FAST
  at Mt. Hopkins.  Compared to earlier epochs \C is about a factor of
  three brighter, allowing us to study the shape of the broad line
  region (BLR) and the dust torus in a high luminosity phase.  { The
    analysis of the light curves yields that the dust echo is rather
    sharp and symmetric in contrast to the more complex broad
    H$\alpha$ BLR echo.  We investigated how far this supports an
    optically thick bowl-shaped BLR and dust torus geometry.
    The comparison with several parameterizations of these models
    supports the following geometry: The BLR clouds lie inside the
    bowl closely above the bowl rim up to a halfcovering angle
    $0^\circ < \theta < 40^\circ$ (measured against the equatorial
    plane).  Then the BLR is spread over many isodelay surfaces,
    yielding a smeared and structured echo as observed.  Furthermore,
    if the BLR clouds shield the bottom of the bowl rim against
    radiation from the nucleus, the hot dust emission comes
    essentially from the top edge of the bowl
    ($40^\circ < \theta < 45^\circ$). Then, for small inclinations as
    for 3C120, the top dust edge forms a ring that largely coincides
    with a narrow range of isodelay surfaces, yielding the observed
    sharp dust echo.  The scale height of the BLR increases with
    radial distance from the black hole (BH).  This leads to luminosity
    dependent foreshortening effects of the lag.  We discuss the
    implications and possible corrections of the foreshortening for
    the BH mass determination and consequences for the lag
    (size) -- luminosity relationships and the difference from
    interferometric torus sizes.   } }

\maketitle

\section{Introduction}
\label{sec:introduction} 

The quasar paradigm comprises a supermassive BH, a
central X-ray source, an accretion disk (AD) surrounded by a broad
line region (BLR), and a molecular dusty torus farther out.  These
formal components serve as working frame: the AD, BLR, and torus may
have smooth transitions rather than being separated entities with
sharp boundaries.  As the inner quasar regions cannot be resolved by
conventional imaging techniques, reverberation mapping (RM) is the
main tool of the trade
(\citealt{1972ApJ...171..467B,1993PASP..105..247P,2004PASP..116..465H}).
The RM technique traces the response of irradiated regions to the
light fluctuations of continuum emission from the inner AD. In
particular, RM studies of the BLR trace the response of line emission
to continuum variations and determine the time lag, $\tau$, between
their signals, with the BLR size, \rblr$\sim c \tau$ ($c$ is the speed
of light).  Reverberation mapping can reach a spatial resolution of
better than $10^{-5}$\,pc, clearly surpassing the capabilities of
current imaging instruments.

Line-to-continuum RM studies find that $R_{\rm BLR}$ ranges from a few
light days to several light months.  At low redshift, a relatively
tight size-luminosity relation for the H$\beta$-emitting region has
been identified: \rblr$({\rm H}\beta)\propto L^\alpha$ with
$\alpha \simeq 0.5$ and monochromatic active galactic nucleus (AGN)
luminosity $L$ at 5100\AA~
(\citealt{2000ApJ...533..631K,2013ApJ...767..149B}).  Conversely, $L$
may be inferred from \rblr or \rdust measurements, opening the
possibility of using AGNs as standard candles for cosmological
distance studies
(\citealt{1999OAP....12...99O,1999MNRAS.302L..24C,2011ApJ...740L..49W,
  2011A&A...535A..73H,2013A&A...556A..97C,2014ApJ...784L..11Y,2017MNRAS.464.1693H}).

To characterize the line profiles, various studies using the ratio
full width at half maximum (FWHM) to line dispersion $\sigma_{v}$
revealed a large diversity between sources with flat-topped lines
(showing large FWHM/$\sigma_{v}$) and sources with narrow-peaked lines
with extended wings (showing small FWHM/$\sigma_{v}$); the latter
class shows a trend toward higher accretion rates as measured by
Eddington ratios, while inclination appears to play a minor role in
FWHM/$\sigma_{v}$ (\citealt{2000ApJ...536L...5S,
  2004A&A...426..797C,2006A&A...456...75C}).  Alternatively, BLR
properties could be luminosity dependent, such that the use of
$\sigma_{v}$ as a proxy for $v_{\rm rot}$ leads to an overestimate of
$M_{\rm BH}$ at high$-z$ or high luminosity.  From RM studies of
individual Seyferts during phases of low and high luminosity
\cite{2011Natur.470..366K,2013A&A...549A.100K,2013A&A...558A..26K}
find an indication that both the scale height above the equatorial
plane and the ratio of turbulent to rotation velocity
$v_{\rm turb} / v_{\rm rot}$ of the BLR increase with luminosity. A
potential explanation is that brightening of an individual AGN makes
the BLR visible at larger $R_{\rm BLR}$, where $v_{\rm rot}$ should be
smaller (Kepler's law); if $v_{\rm turb}$ remains constant then the
ratio $v_{\rm turb} / v_{\rm rot}$ increases with $R_{\rm BLR}$.

If the height-to-radius ratio $H/R $ is proportional to
$ v_{\rm turb} / v_{\rm rot}$, this implies a growing vertical extent
of the BLR above the disk plane.  It is clear that geometry has a
substantial influence on the calculation of the central BH
mass.

Well-sampled light curves allow us to restore information about the
spatially unresolved BLR geometry
(\citealt{1991ApJ...367L...5H,1991ApJ...367..493M,2012ApJ...754...49P}).
In a nutshell, tightly localized gas distributions (e.g., nearly
face-on flat disks) produce a sharp echo of the continuum light curve,
while more isotropic gas configurations (e.g., spheres) lead to
smoothed light curves for the line emission band.  The line-integrated
transfer function depends on the BLR geometry, i.e., the location of
the BLR clouds at the time they respond; BLR kinematic information is
not necessary.  It is intriguing to check if the geometry of the
visible BLR changes with luminosity (or accretion rate \mdot) in a
nonisomorphic manner.

For the archetypal Sy-1 galaxy NGC\,5548, \cite{2007arXiv0711.1025G} analyzed the AGN energy budget and derived crucial conclusions concerning
the geometry of the BLR and dust torus: the BLR has a likely
covering factor about 40\%, which translates to a half-covering angle
$\theta \approx 40^{\circ}$.  The BLR shields a substantial fraction
of the dust torus from direct illumination by the AD, allowing for the
observed relatively small near-infrared (NIR) contribution to the AGN energy
budget. At the same time, the BLR obscuration also removes the problem
that the dust torus covering factor is greater than the BLR covering
factor, and is consistent with the observed fraction of obscured
AGNs.  The flux reduction at the torus also reduces the problem of AGN
dust reverberation lags giving sizes smaller than the dust sublimation
radii.

Near-infrared RM studies of the dusty torus find
$\tau_{\rm dust} \approx 4 \times \tau_{H\beta}$ between hot-dust
continuum emission and optical continuum fluctuations
(\citealt{2014ApJ...788..159K}).  However,
$R_{\rm dust} = c \cdot \tau_{\rm dust}$ is three times smaller than
the dust sublimation radius, $R_{sub}$, inferred from the UV
luminosity (\citealt{2006ApJ...639...46S,2007A&A...476..713K}).  {To
  resolve this conflict, \cite{2010ApJ...724L.183K,
    2011ApJ...737..105K} proposed a bowl-shaped dust torus that
  smoothly continues into the central AD.  The AD
  emission is highest at the polar region and lowest at the equatorial
  region.  The anisotropy of the AD emission controls the angle
  dependent dust sublimation radius and thus the concave rim of the
  bowl.  For the bowl-shaped torus, assuming that the hot dust
  emission arises from the entire bowl surface, the dust transfer
  functions show about three times faster responses than for a
  bagel-shaped torus (see, e.g., Fig.~1 of
  \citealt{1995PASP..107..803U}), in agreement with observations.
  Notably, for small inclinations (i.e., rather face-on than edge-on
  view) the transfer functions of the dust emission show a sharp peak.
  Kawaguchi \& Mori's studies pioneered the bowl-shaped dust torus,
  but did not yet consider the BLR.  The relation between BLR and dust
  torus has long been debated.  For instance,
  \cite{2011A&A...525L...8C} proposed that the AD reaches far out
  until it meets the torus and that the BLR clouds are launched from
  the outer AD in a dusty wind at that radius, where the temperature
  in the AD matches the dust sublimation temperature. In this scenario,
  however, how far the reverberation-based BLR and
  dust lags should essentially be the same is an open issue contrary to the factor 3-5
  larger dust lags observed so far.

  Consequently, in continuation of Kawaguchi \& Mori's work,
  \cite{2012MNRAS.426.3086G} modeled BLRs, assuming that they are
  confined by a bowl-shaped torus geometry, and suggested that the BLR
  clouds lie above the bowl surface (see their Fig.~1).  Goad et
  al. successfully tested the bowl-confined BLR models with
  reverberation data of NGC\,5548.  Notably, the BLR clouds may shield
  part of the dust from the AD radiation, bringing the hot dust
  covering fraction closer to that deduced by
  \cite{2011MNRAS.414..218L} and \cite{2011ApJ...737L..36M}.  Then the
  hot dust emission would arise essentially from the edge, i.e., top
  rim, of the bowl.  For the Seyfert-1 WPVS48,
  \cite{2014A&A...561L...8P} found an exceptionally sharp NIR echo,
  which led these authors to favor the theory that the hot dust is
  essentially located at the edge rather than the entire rim.

  Simultaneous BLR and dust reverberation studies might be able 
  to shed further light on the geometry and the interplay between BLR and dust.
  Therefore, we have embarked on such a study of 3C\,120.
}

The BLR galaxy \C belongs to the most frequently
monitored Seyfert-1 galaxies and results from several campaigns have
been reported:\ spectroscopic RM by
\cite{1998PASP..110..660P} covering the H$\beta$ region, velocity
resolved RM covering H$\gamma$, H$\beta,$ and
H$\alpha$ by \cite{2012ApJ...755...60G,2013ApJ...764...47G} and
\cite{2014A&A...566A.106K}.  Photometric reverberation mapping {(PRM)}
of H$\beta$ has been reported by
\cite{2012A&A...545A..84P,2014A&A...568A..36P}.  During the
spectroscopic campaigns before 1998 and in 2008 and 2009 (Kollatschny) the
average $\lambda log(L_{\lambda 5100})$ AGN luminosity at 5100\AA\ was
about ${44.01}$ and ${44.12}$.  In 2009 and 2010 (Pozo Nu\~nez) this
dropped to ${43.84}$\,erg\,s$^{-1}$, similar as in 2010 and 2011 (Grier,
${43.87}$\,erg\,s$^{-1}$).  Between September 2014 and March 2015 we
continued photometric monitoring; \cite{2015A&A...581A..93R} found a brightening by a factor
three (${44.32}$\,erg\,s$^{-1}$) from the $B$ and $V$ band light
curves.  The brightening occurred between
January and August 2014, as revealed by a spectrum in December 2013
taken at the Asiago Observatory where
$\lambda log(L_{\lambda 5100}) \approx 43.82$\,erg\,s$^{-1}$ (PhD
thesis, private communication, data in \citealt{2015A&A...578A..28B}).

We report on the analysis of the entire data set in $BVRIJK$ and
a narrow band (NB) at 680\,nm and contemporaneous spectrum taken in
November 2014. We assumed a luminosity distance of 138\,Mpc for a
cosmological model with $\Omega_m = 0.27$, $\Omega_v = 0.73$ and
$H_0 = 73$\,km\,s$^{-1}$Mpc$^{-1}$.

\section{Observations and data reduction}
\label{sec:Observations}
We monitored \C between 27 August 2014 and 3 March 2015 at the \USB, which is located near Cerro Armazones, the
location of the upcoming ESO Extremely Large Telescope.

The photometric data was reduced with standard IRAF\footnote{{ IRAF is
    distributed by the National Optical Astronomy Observatory, which
    is operated by the Association of Universities for Research in
    Astronomy (AURA) under cooperative agreement with the National
    Science Foundation.}}
bias, dark, and flatfield correction. Astrometric matching was
performed with \SCAMP\ (\citealt{2006ASPC..351..112B}). Before
stacking multiple exposures, they were resampled onto a common
coordinate grid using \SWARP (\citealt{2002ASPC..281..228B}). The
photometry is performed on combined frames with a fixed $7\farcs5$
aperture, found to be the optimum in our previous studies
(\citealt{2011A&A...535A..73H,2012A&A...545A..84P,2013A&A...552A...1P,2015A&A...576A..73P,2015A&A...581A..93R}).

Our aim is to combine relative photometry in the field with absolute
photometry anchored to Landolt fields. First, we selected 20 close
nonvariable stars (calibrators) in the AGN field with brightnesses
ranging from 0.5 to 10 times of the nuclear \C flux. Then, we performed photometry
of \C and these stars using the $7\farcs5$ aperture with
SExtractor (\citealt{1996A&AS..117..393B}). The calibrator light curves
were normalized to their average over all days in the observation
epoch. We used the scatter of the normalized ensemble of calibrators within a
particular night as the relative photometric error. Dividing the
flux of \C by the nightly average of the normalized ensemble resulted
in the relative light curve of the target.

Absolute photometry was obtained using additional observations of
suitable fields from \cite{2009AJ....137.4186L}. Since there are no
literature photometric fluxes for the NB and $R$ bands\footnote{We do
  not use the Cousins filter system $R_c$. The $\lambda_{\rm eff}$ of
  our $R$ band lies at 700\,nm.}, we fit the available Landolt
measurements of each source by Planck functions. We rejected the star for
calibration if the largest
residual of this fit was larger than 5\%. Otherwise, the flux of the NB was interpolated by the
Planck function. We performed airmass dependent extinction correction (\citealt{2011A&A...527A..91P}) to determine the instrument response
in the \C field in units of Jy. Finally, galactic foreground
extinction correction was applied based on the
\cite{2011ApJ...737..103S} extinction maps. The calibration of the
relative IR fluxes in $J$ and $K$ band was performed in the same manner
as for the optical data above. We carried out the absolute calibration using 2MASS stars (AAA flagged) in the \C maps.
Supplementing our photometric data, we obtained a single epoch
spectrum taken by the SAO FAST spectrograph at the Fred Lawrence
Whipple Observatory (FLWO) on Mt. Hopkins.

\subsection{Berlin Exoplanet Search Telescope II}
\label{ssec:best}

The Berlin Exoplanet Search Telescope II (BEST\,II) is a 25\,cm
aperture Baker-Ritchey-Chr\'{e}tien system, using a KAF16801
$4096\times4096$ pixel CCD with a pixel size of 9\,$\mu$m and a field
of view of $1.7^{\circ}\times1.7^{\circ}$. Since 2012, this telescope is also in
service for AGN observations. More details on the instrument can be
found in \cite{2009A&A...506..569K}.

The \BII telescope was used with Johnson $B$, $V$, $R$
($\lambda_{\rm eff}=700$\,nm) and $I$ bands to trace the
AD continuum variations. This allows us to separate the
host galaxy flux in these filters (see \citealt{2015A&A...581A..93R})
and to remove the AGN continuum contribution in our NB.

\subsection{Robotic Bochum Twin Telescope}
\label{ssec:robott}

The 15\,cm Robotic Bochum Twin Telescope (RoBoTT), previously known as
the VYSOS\,6 telescope, is a twin-refractor design. It consists of two
Takahashi TOA 150, attached to a common German equatorial mount. Both
refractors are equipped with KAF16801 CCDs, hence they provide the
same quantum efficiency per wavelength as the \BII telescope when
using identical filters. The field of view is
$2.7^{\circ}\times2.7^{\circ}$.  More details can be found in
\cite{2012AN....333..706H}.

\Vy6 Johnson $V$ band observations serve to supplement the \BII data
by covering a gap between September and October 2014. Before and after
this gap, there are numerous days at which both instruments were
available to observe simultaneously. The overlap between \BII and \Vy6
in Fig.  \ref{fig:lc_2014} demonstrates the comparability of both
instruments.

\subsection{Bochum Monitoring Telescope}
\label{ssec:bmt}

The Bochum Monitoring Telescope (\V16) is a 40\,cm Coud\'{e}
telescope, featuring a SBIG STL-6303 CCD with $3072\times2048$ pixel,
each sized 9\,$\mu$m. The resulting field of view is
$41.2' \times 27.5'$. A more detailed description of the telescope and
filters can be found in \cite{2013AN....334.1115R}.

To monitor the redshifted \Ha emission line, NB
observations in an Asahi 680\,nm filter (12\,nm FWHM) have been
provided.  Owing to the redshift of $0.03301 \pm 0.00003$
(\citealt{1988PASP..100.1423M}), the broad \Ha emission line is
redshifted into our NB filter, centered at $677.94$\,nm.

\subsection{Infra-Red Imaging System}
\label{ssec:iris}

The Infra-Red Imaging System (IRIS) is a 80\,cm Nasmyth telescope
equipped with a HAWAII-1 detector.  The field of view is
$12.5' \times 12.5'$ with a pixel size of
$0\farcs74 \times 0\farcs74$.  The filter wheel is equipped with
standard 2MASS $J,H,K,$ and NB filters.  Further information
can be found in \cite{2010SPIE.7735E..1AH}.
\subsection{FAST Spectra}
\label{ssec:fast}

\begin{figure}
  \includegraphics[angle=0,width=\columnwidth]{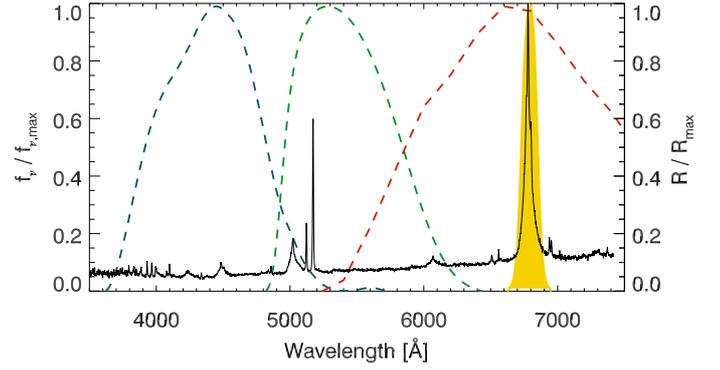}

  \caption{The FAST spectrum of November 21$^{\rm st}$ 2014 is
    normalized to the maximum value and drawn as black
    line. Overplotted in colored dashed curves are the Johnson $B,\,V$
    and $R$ responses that have been folded with the KAF 16801
    CCD. The yellow shaded area represents the 680\,nm NB filter
    folded with the quantum efficiency of the STL-6303 CCD. All filter
    response functions are normalized to their maximum.}

  \label{fig:spec}
\end{figure}

To supplement our new photometric data of 2014/2015 we used the FAST
long-slit spectrograph of the 1.5\,m Tillinghast reflector telescope
at the Fred Lawrence Whipple Observatory in Arizona, USA. Spectra of
\C together with the calibration star HD\,19445 were taken in the
night of 21 November 2014. The 300 lines\,mm$^{-1}$ grating
was used together with a $3\farcs0$ slit aperture. More details on the
instrument are given in \cite{1998PASP..110...79F}. Reduction was
performed using standard flatfield, wavelength, and flux calibration
routines as described in \cite{1997ASPC..125..140T}.  The resulting
spectrum 
is shown in Fig.~\ref{fig:spec} together with the filter response
profiles of our photometric campaign.

\section{Results}
\label{sec:results}

\subsection{Line coverage of the narrow band}
\label{ssec:line_contrib}

The FAST spectrum (Fig. \ref{fig:spec}) allows us to estimate the
contribution of the continuum and the narrow and broad emission lines
to the photometric bands. For this purpose, we fold the filter
response curves with the quantum efficiency of the KAF-16801 CCD and
use this as detector response function $R(\lambda)$.  For each
selected band we integrate the line profiles $L(\lambda)$ and the
(power law) continuum $C(\lambda)$ separately. The fraction $f$ of
line flux of the total flux is then given in
Eq.~\ref{eq:filtercontrib} as follows:

\begin{equation}
  \label{eq:filtercontrib}
       f = \frac{ \int \! R(\lambda) L(\lambda)  \, \mathrm{d}\lambda}{\int \! R(\lambda)  \left[ L(\lambda)+C(\lambda) \right]  \, \mathrm{d}\lambda}
.\end{equation}
The results of the line contributions are summarized in Table
\ref{tab:spec}.  The contributions of H$\gamma$ and \Hb to the $B$
band flux are on the order of our photometric uncertainty. The
contribution of \Hb to the $V$ band is only slightly above this
value. The flux beyond $7400$\,\AA\ was assumed to be the extrapolated
power law, determined from the continuum. In the $R$ band, the strong
\Ha emission causes a contribution of $18\%$.  The $680$\,nm NB filter
is clearly dominated by the \Ha flux.

\begin{table}
  \begin{center}
    \hfill{}
    \caption{Contribution of emission lines to the photometric
      bands. The percentages were obtained from the spectrum by
      integrating the emission lines convolved with the filter
      responses of the telescopes.}
    \label{tab:spec}
    \begin{tabular}{@{}l|cccc}
      \hline
      \hline
      Band & \Ha & \Hb & H$\gamma$& OIII \\
           & $\%$ &  $\%$ &  $\%$ &  $\%$ \\
      \hline
      $B$ &  -- & $2.8$ & $3.6$& -- \\
      $V$ &  -- & $5.0$ & -- & $7.3$ \\
      $R$ &  $18$ & -- & -- & -- \\
      $680$\,nm&  $71$ & -- & -- & -- \\
    \end{tabular}
  \end{center}
\end{table}

\begin{figure}
  \includegraphics[angle=0,width=\columnwidth]{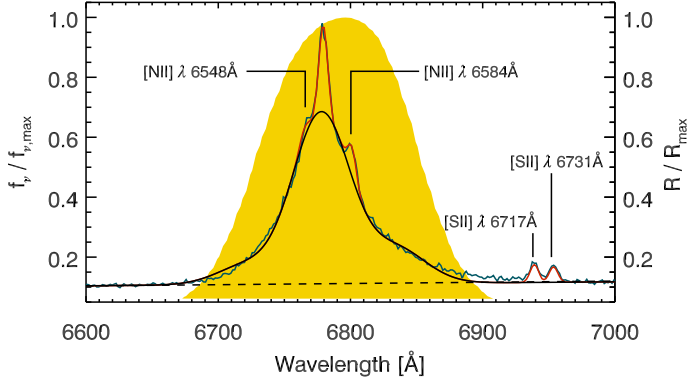}
  \caption{\Ha line profile decomposition. The continuum is shown with
    a black dashed line. The narrow \Ha, [N\,II], and [S\,II] features
    were fitted with the red curves. The yellow area shows the
    680\,nm NB profile. All data is normalized to the
    maximum. The broad \Ha is represented by the black continuous
    curve.}
  \label{fig:spec_moment}
\end{figure}

While spectroscopic reverberation data enables us to use the velocity
information from the RMS spectra, i.e., from the variable portion of
the emission lines, several attempts have been made to determine the
mass of the central BH using single epoch spectra (e.g.,
\citealt{2002ApJ...571..733V}, \citealt{2007ApJ...661...60W},
\citealt{2009ApJ...692..246D}).

In order to calculate the velocity dispersion of the broad \Ha line,
we perform a decomposition of the spectrum around this feature into
the following components:

\begin{enumerate}

\item As depicted by the dashed black line in
  Fig. \ref{fig:spec_moment}, we assume the continuum in the
  wavelength range from $6600\,\AA$ to $7000\,\AA$ to follow a power
  law (it is almost linear in this range). The power law was
  determined through fitting in the line free spectral regions
  $4600-4800,$ $5300-6000,$ $6300-6500,$ and $7050-7200$\,\AA.

\item The narrow \Ha, [N\,II], and [S\,II] lines are fitted through
  subtraction of a line profile template of the strong $5007\,\AA$
  [O\,III] emission line, positioned at the appropriate redshifted
  wavelength.  To improve the fitting quality of the faint $6548\,\AA$
  [N\,II] component, we assume its integrated flux to be 1/3.05 of the
  $6584\,\AA$ [N\,II] line, according to line intensity ratios as
  determined by \cite{2000MNRAS.312..813S}.  While it is usually more
  reasonable to use a template from the [S\,II] lines
  (\citealt{1997ApJS..112..391H}, due to the similar critical density
  as [N\,II]), these lines in our spectrum are weak and blended, so we
  approximate the shape of the narrow lines by the [O\,III] profile
  instead.  The ratio of best fit narrow \Ha to [N\,II] $6584\,\AA$
  line strength is $2.96/1$.  These profiles are shown by the red
  curves in Fig. \ref{fig:spec_moment}.

\item For the remaining broad \Ha line, we use a decomposition into
  Gauss-Hermite polynomials similar to
  \cite{2007ApJ...661...60W}. Using a fourth order polynomial produces
  the curve shown by the black solid line in
  Fig. \ref{fig:spec_moment}.  { Fitting the broad line profile with
    two or more Gaussians does not deliver a better fit than with
    Gauss-Hermite polynomials.  }

\end{enumerate}
Correcting for the spectral resolution of the FAST instrument, we
determine the broad \Ha dispersion of \C\ to be
$1532 \pm201$\,km\,s$^{-1}$.  This is lower but still comparable to
the result of $1638\pm105$\,km\,s$^{-1}$, by
\citealt{2014A&A...566A.106K} using the Hobby-Eberly Telescope for
spectroscopic RM in a different epoch (2008/2009). The FWHM of the
{broad} \Ha\ {component alone} is $1420\pm129$\,km\,s$^{-1}$.

\subsection{Light curves}
\label{sec:lightcurves}

\begin{figure}
  \centering
  \includegraphics[width=\columnwidth]{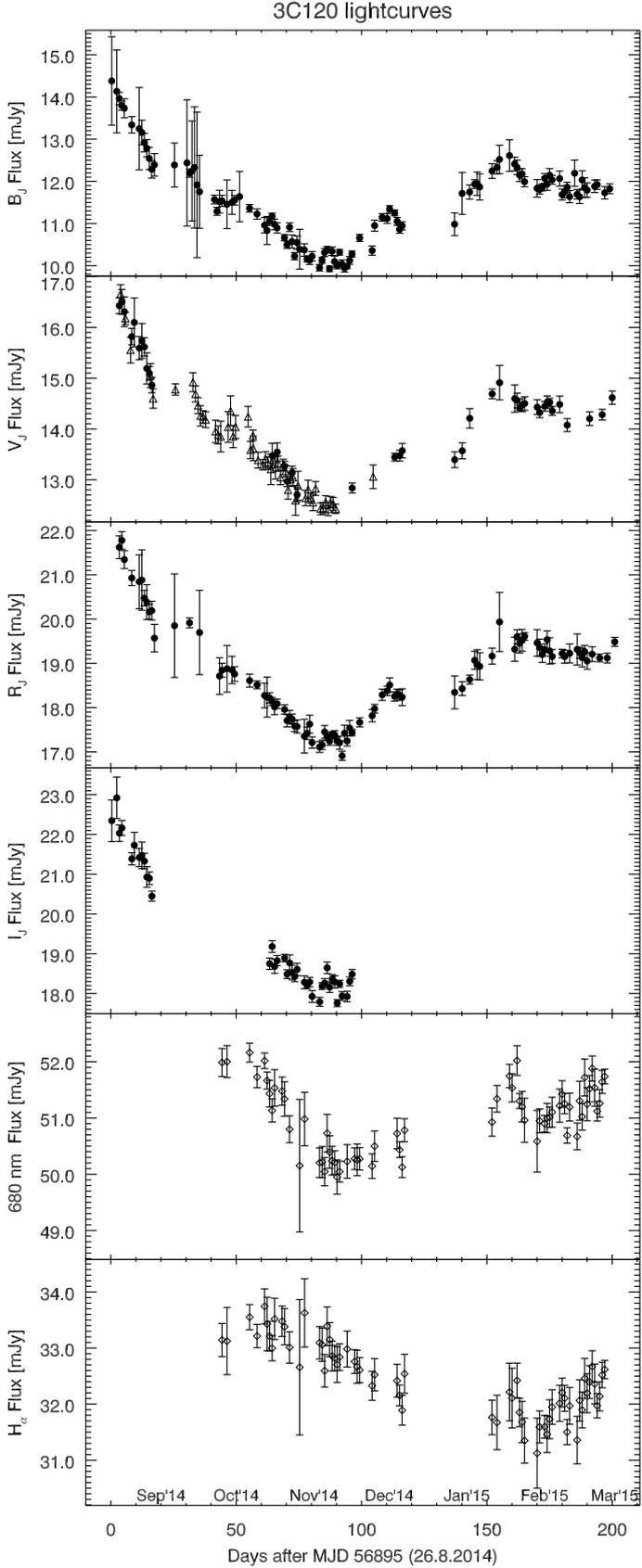}
  \caption{Combined optical light curves from all telescopes.  Open
    diamonds indicate observations taken with the \V16 telescope, filled
    circles represent \BII and open triangles \Vy6. All fluxes were
    measured inside an aperture of $7\farcs5$ and corrected for
    galactic foreground extinction. The \Ha light curve (680\,nm - $R$
    band) is shown at the bottom.}
  \label{fig:lc_2014}
\end{figure}
The optical light curves in Fig. \ref{fig:lc_2014} show significant
variability features in all bands. The $B,\,V,$ and $R$ band
observations, tracing the variable continuum, start with a decline in
brightness from the absolute maximum values in September 2014 to the
minimum in mid-November 2014. Afterward there is a rise of flux to a
local maximum in mid-December 2014. Between mid-December and early
January 2015, there appears to be a slight decline that changes into
another rise toward end of January.  Afterward, we observe a plateau
phase with minor variations. The 680\,nm NB flux already reveals the
large contribution of \Ha emission. This flux follows the behavior of the
broad bands with a distinct lag, but the relative amplitude of the
variations is much lower. The NB contains in addition to the strong
\Ha  a substantial continuum contribution that has to be removed
for a proper lag determination.

The \Ha light curve was obtained by subtracting the continuum flux from
the 680\,nm NB as described in Section \ref{ssec:separation}.  \Ha
clearly shows similar features as the broad optical bands but shifted
in time. However, compared to the average flux, the amplitude of these
variations is much lower and is analyzed in Section \ref{ssec:dcf}.

Observations in $I$ band between September and December 2014 trace the
brightest and faintest phases, and therefore are valuable for the flux
variation gradient (FVG) analysis.

\subsection{Flux separation}
\label{ssec:separation}

\subsubsection{AGN and host in the optical}
\label{sssec:agn_host}

Using the FVG method
(\citealt{1981AcA....31..293C,1992MNRAS.257..659W}), we disentangle
the constant host from the variable AGN flux inside our aperture. This
is possible since the color of the variable component is sufficiently
constant.  The method was already applied successfully to the $B$ and
$V$ band data of \C, covering observations from 2009 until 2015 in
\cite{2015A&A...581A..93R}. Owing to the large luminosity coverage of
the underlying $B$ and $V$ band data, we adopt the determined $B$ band
host flux of $(1.69 \pm 0.28)$\,mJy and the corresponding $V$ band
host flux of $(3.89 \pm 0.29)$\,mJy.

\begin{figure} \centering
  \includegraphics[width=\columnwidth,clip=true]{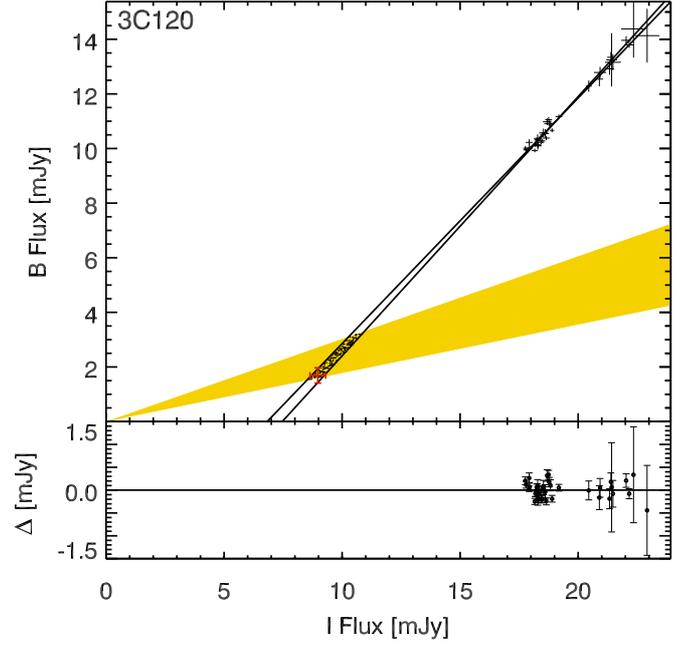}

  \caption{$B$ vs. $I$ flux variations of \C in the $7\farcs5$
    aperture and the residuals of the data to the linear regression.
    Each point is drawn as a thin cross in which the line length
    corresponds to the photometric uncertainties in the respective
    filters.  The yellow cone denotes the host color range of
    \cite{2010ApJ...711..461S}.  The black continuous line covers the
    upper and lower standard deviations in the AGN slope
    $\Gamma_{BI}$, given by the OLS bisector fit. The intersection of
    $B$ host flux and AGN slope is indicated by the red cross. }

  \label{fig:fvg_bi}
\end{figure}

The contribution in the $I$ band is determined through the FVG diagram
in Fig.  \ref{fig:fvg_bi}.  The continuous black lines indicate the upper
and lower 1$\sigma$ deviation of the flux variation of the AGN with a
slope $\Gamma_{BI}$ of $0.925\pm0.025$ determined by an ordinary least
square (OLS) bisector fit. The intersection of the AGN and host flux
of $B=(1.69\pm0.28)$\,mJy occurs at $I=(8.97 \pm 0.33)$\,mJy, which is
denoted by the red cross. The mean AGN flux in the $I$ band is
determined as the average of the remaining variable component, which
is $(10.34 \pm 1.54)$\,mJy. All fluxes are listed in Table
\ref{tab:tau_fluxes}.

For comparison, the yellow filled area outlines the range of host
$B$-$V$ colors as determined for several Seyfert 1 galaxies by
\cite{2010ApJ...711..461S} in an $8\farcs3$ aperture. Using the
$B$-band flux of our long-term campaign
(\citealt{2015A&A...581A..93R}), our value lies at the lower (red)
range.

\begin{figure}
  \centering
  \includegraphics[width=\columnwidth,clip=true]{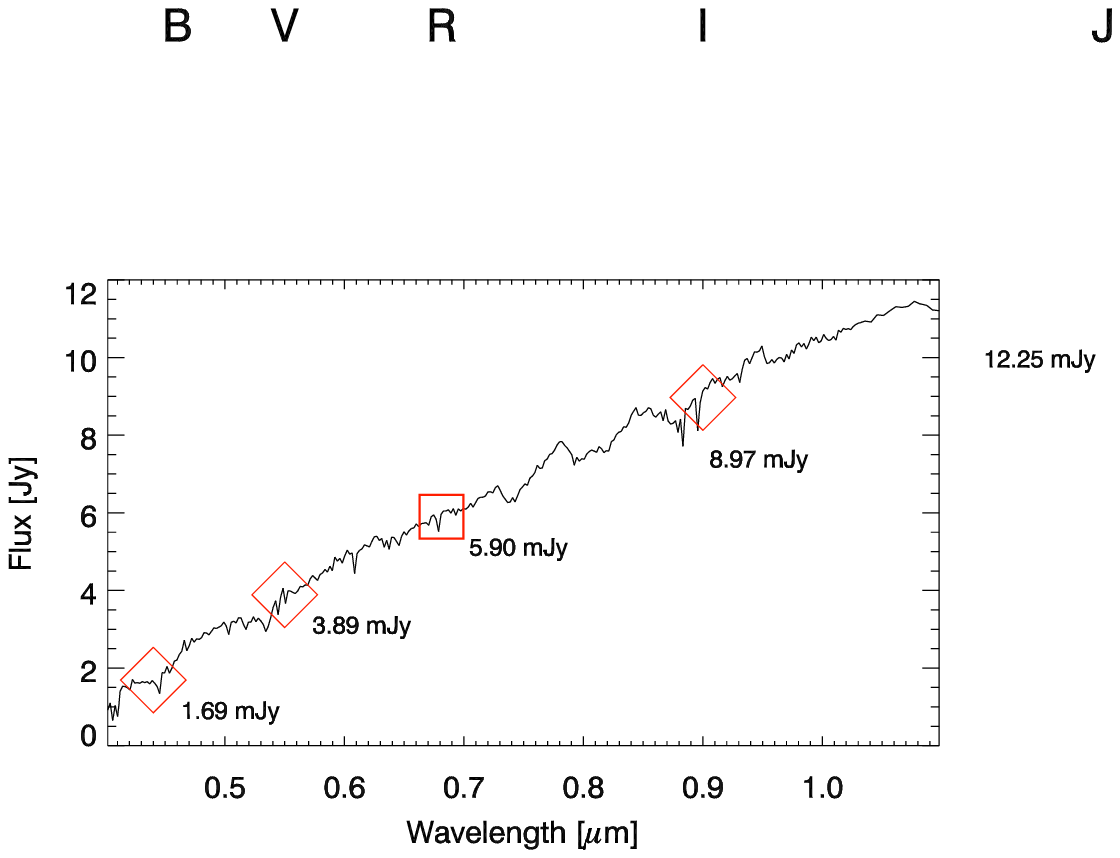}
  
  \caption{Host galaxy template spectrum in the observed filters,
    scaled to match the measured fluxes in $B,V,$ and $I$ (red
    diamonds). The red square shows the fitted $R$-band flux.} 

  \label{fig:host_interpolated}
\end{figure}

Through image decomposition of HST images, \cite{2009ApJ...697..160B}
found an S0 host morphology for 3C\,120. To estimate the host fluxes
in the $R$ band, we fit the $BVI$ spectral energy distribution (SED)
with a redshifted S0 galaxy template. The best fitting spectral
template was obtained with the PEGASE-3 spectral evolution modeling
package for an age of 13\,Gyr (Fioc et al. 2017, in prep., based on
\citealt{1997ASSL..210..257F}). Fig. \ref{fig:host_interpolated} shows
the optical to infrared part of the spectrum. The spectral template is
redshifted to $z=0.03301$ and flux-scaled to our measured $B,V,$ and
$I$ host fluxes indicated by the diamond symbols. Fluxes are
determined by convolution of the PEGASE3 spectrum with the response
function of our filters. Assuming that this template is a good
approximation, the expected $R$-band flux of the host is then
$( 5.90 \pm 0.31)$\,mJy, shown by the red square in
Fig.~\ref{fig:host_interpolated}.

\subsubsection{\Ha emission line flux}
\label{sssec:ha_separation}

\begin{figure}
  \centering
  \includegraphics[width=\columnwidth,clip=true]{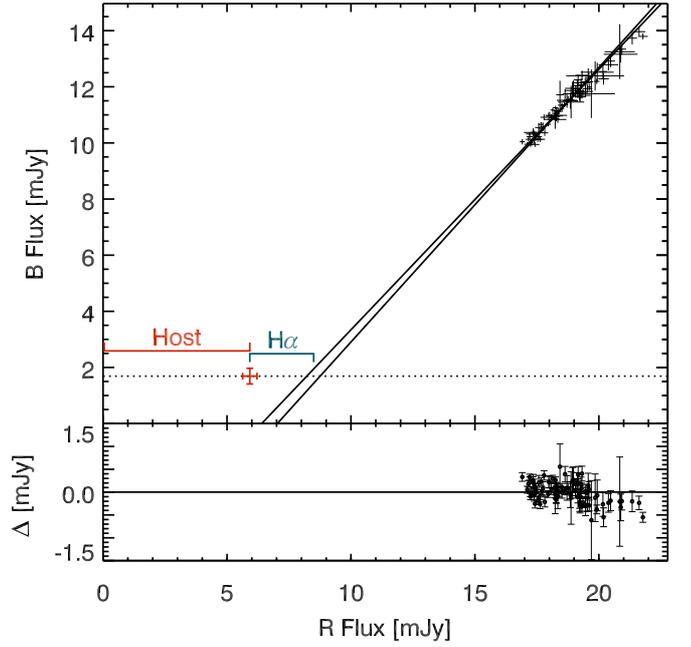}

  \caption{$B$ vs. $R$ flux variations of \C in the $7\farcs5$
    aperture. Same notation as in Fig. \ref{fig:fvg_bi}. The dotted
    black line follows the $B$-band host flux. }

  \label{fig:fvg_br}
\end{figure}

With the estimate of host flux in the $R$ band, we can now take a
closer look at the flux variation behavior seen in the $B-R$ FVG
diagram in Fig. \ref{fig:fvg_br}. We interpolate the expected slope of
the AGN to be $\Gamma_{BR}=0.953\pm0.013$ using
$\Gamma_{BV}=0.979\pm0.005$ (\citealt{2015A&A...581A..93R}) and
$\Gamma_{BI}=0.925\pm0.025$ as determined above and assuming a
power law AGN SED between $B$ and $I$. This slope is now aligned to
the observed fluxes in the $B$ and $R$ bands. As seen in the residuals
in Fig. \ref{fig:fvg_br}, the flux variations of the continuum and
\Ha, as seen in the $R$ band, show a very linear trend that suits our
interpolated slope $\Gamma_{BR}$ within the majority of error
margins. Some deviation of linearity may occur since the $R$ band
contains the delayed \Ha variation that does not correlate directly
with the variable continuum. However, the relative amplitude of the
\Ha variation is small ($<10$\,\%), as seen in
Fig.~\ref{fig:lc_2014}. The $R$ band host flux is indicated by a red
cross in Fig.~\ref{fig:fvg_br}. Clearly, the AGN slope does not
intersect our measured $B$-band flux (dotted line) at this
position. This is due to the contribution of the \Ha emission line in
the $R$ band. The average \Ha flux can then be determined from this
FVG diagram considering the constant $R$-band host flux estimate of
$5.90$\,mJy (Sect.~\ref{sssec:agn_host}). Then the average \Ha
flux\footnote{Including the NB emission lines as in
  Fig. \ref{fig:spec_moment}.}  in the $R$ band is
$(2.57\pm0.42)$\,mJy. The corresponding flux contribution of \Ha in
the $R$ band is then $(13.7\pm2.2)\,\%$. This is less than estimated
from the smaller slit aperture of the spectroscopic data (see
Tab.~\ref{tab:spec}). Becasue of the smaller aperture, more flux of the
spatially extended host is lost in the spectrum.

Because the central wavelength of the NB filter lies close to
$\lambda_{\rm eff}$ of the $R$ band, we can assume that the host to
AGN ratio in the continuum is roughly the same in both filters. This
allows us to subtract the $R$-band light curve directly from the NB
light curve in order to obtain the variable signal of \Ha alone. The
resulting \Ha light curve is depicted at the bottom of
Fig. \ref{fig:lc_2014}, showing a minimum in Feb. 2015, and hence a
clear delay relative to the AGN continuum variation seen in the broad
bands.

\subsection{Delay of the \Ha broad line region}
\label{ssec:dcf}

The large three month decline and two month rise is seen in all the broad
bands.  It is followed by a plateau phase during which the flux is
almost constant, close to the average flux of the epoch. Any long-term
secular variation is masked within the structure of the long-term
feature. Its response can be clearly seen in the \Ha light curve. The
shift of this signal is about 70 days, which is about half of our
total observation campaign. Linear detrending as proposed by
\cite{1999PASP..111.1347W} would cause a bias toward short
delays and is therefore not performed.
\begin{figure}
  \centering
  \includegraphics[width=\columnwidth,clip=true]{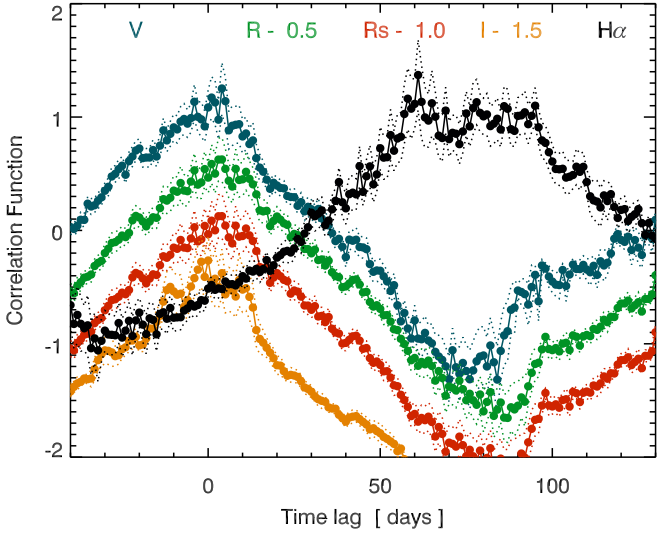}

  \caption{Discrete correlation functions between $B$ and the optical
    data. The correlation of $B$ with $V$ band is colored blue; the
    uncorrected $R$-band correlation is shifted -0.5 and shown in
    green; the line flux corrected $R$ band is shifted -1.0 and shown
    in red; the $I$-band correlation is shifted -1.5 and shown in
    orange; the correlation between $B$ and \Ha is shown in
    black. Dotted lines represent the 1$\sigma$ uncertainties.}

  \label{fig:dcf}
\end{figure}
    
\begin{figure} \centering
  \includegraphics[width=\columnwidth,clip=true]{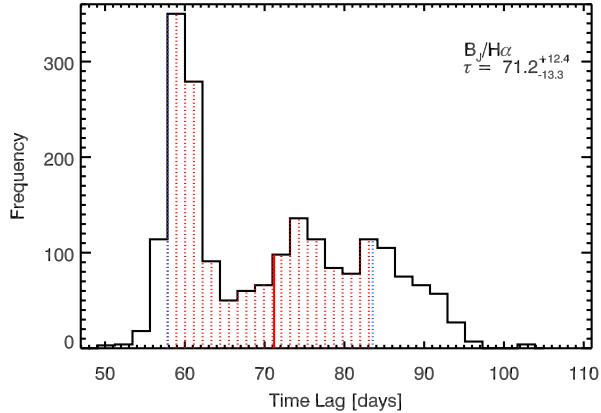}
  \caption{ $B-H\alpha$ cross correlation (DCF) statistics obtained
    from 2000 FR/RSS runs.  The vertical red solid line denotes the
    median.  The vertical red dotted area indicates the range inside the
    68\% confidence, and the blue vertical dotted lines give the
    formal 1$\sigma$ uncertainty around the median.  We note the
    low-$\tau$ peak, i.e., the strong signal at lag 60 days, about 20\%
    shorthand of the median.  }
  \label{fig:dcf_BHa}
\end{figure}

To determine the \Ha delay, we apply the discrete correlation function
(DCF; \citealt{1988ApJ...333..646E}) analysis between $B$, \Ha and
other optical light curves. A binning of one day is chosen since this
represents the median sampling of our data set.  Fig.~\ref{fig:dcf}
shows for $B$-\Ha a clear broad maximum ranging from 55 to 95 days.
To estimate the uncertainty, the flux randomization/random subset
selection (FR/RSS) method is used as described by
\cite{1998PASP..110..660P}. This leads to a result of
$\tau_{\rm obs}=71.2^{+12.4}_{-13.3}$\,days for the delay of \Ha in
the observers frame. All delays are finally corrected for redshift
dependent time dilation and listed in Table~\ref{tab:tau_fluxes}.

Notably, the width of the cross correlation (from 55 to 95 days,
i.e., 40 days, Fig.~\ref{fig:dcf}) is significantly larger than
indicated by the $\tau_{\rm obs}$ uncertainty of 1-$\sigma$ $\approx$
13 days derived by the FR/RSS method.  This plateau-like cross
correlation indicates even a substructure in the form of a sharp bright
peak at 60 days.  Fig.~\ref{fig:dcf_BHa} shows the histogram of the
FR/RSS statistics in more detail.  It clearly reveals a sharp peak
around 60 days, in addition to the broad plateau between 55 and 95
days.  This suggests a complex transfer function, if the input
variation is a delta function.
     
\subsection{Inter-band continuum delays}
\label{ssec:dcf_cont}

The dense (one day) median sampling of the data allows us to search for
delays that increase with increasing wavelength and are associated with
reprocessing of emission in the AD, as was reported for
several other Seyfert 1s by \cite{2005ApJ...622..129S}. Compared to
the measured \Ha and \Hb BLR sizes of 3C\,120 of several light weeks,
the size of the AD is expected to be much smaller, on the
order of light days.

As seen in the spectrum (Fig. \ref{fig:spec}), the $V$ band is slightly
polluted by \Hb while \Ha has $18$\% contribution in the $R$ band. Our
photometric analysis (Sect. \ref{ssec:dcf}) shows that the
contribution of \Ha to $R$ in the photometry is even weaker ($13.7$\%
in $R$) and that the amplitude of the line variation is much smaller
than that of the continua. To verify that the \Ha contribution to the
broad bands does not introduce a significant bias toward larger
delays, we also compute an $Rs$ band light curve from our $R$ band
minus the \Ha light curve.

The correlation functions between $B$, $R$, $Rs$, and $I$ are shown in
Fig. \ref{fig:dcf}.  The results of the FR/RSS analysis between $B$
band and other continuum light curves are summarized in Table
\ref{tab:tau_fluxes}.  The errors are consistent with a very low delay
close to $\tau \approx 0$. A bias toward higher delays, potentially
introduced by the \Hb contribution, is not seen.

\begin{table}
  \begin{center}
    \hfill{}
    \caption{Mean fluxes from the FVG analysis and rest-frame
      delays. All delays are relative to the $B$ band. Round brackets
      enclose extrapolated values (Fig. \ref{fig:fvg_br}).  }
    \label{tab:tau_fluxes}
    {\renewcommand{\arraystretch}{1.5}
      \begin{tabular}{@{}l|ccc}
        \hline
        \hline
        Filter&     $\tau$  & Accretion disk  & Host  \\
           &      [days] & [mJy] & [mJy] \\
           \hline
           $B$   &  --                &$9.86\pm1.12$  &$1.69\pm0.28$ \\
           $V$   &$-1.1^{+1.6}_{-3.8}$&$10.06\pm1.18$ &$3.89\pm0.29$  \\
           $R$   &$1.6^{+1.6}_{-3.2}$ &$10.22\pm1.12$ &($5.90\pm 0.31$)\\
           $Rs$ &$1.6^{+1.5}_{-4.0}$ & --            & --             \\
           $J$   &$93.0^{+2.6}_{-6.6}$& -- & -- \\
           $K$   &$94.4^{+4.1}_{-6.9}$& -- & --\\
           $680$\,nm &$21.6^{+15.6}_{-15.7}$& --     & --            \\
           H${\alpha}$&$68.9^{+12.4}_{-13.3}$ &   -- & --             \\

      \end{tabular}
    }
  \end{center}
\end{table}

\subsection{Dust torus variability}
\label{ssec:torus_var}
\begin{figure} \centering
  \hspace{-.15cm}\includegraphics[width=\columnwidth,clip=true]{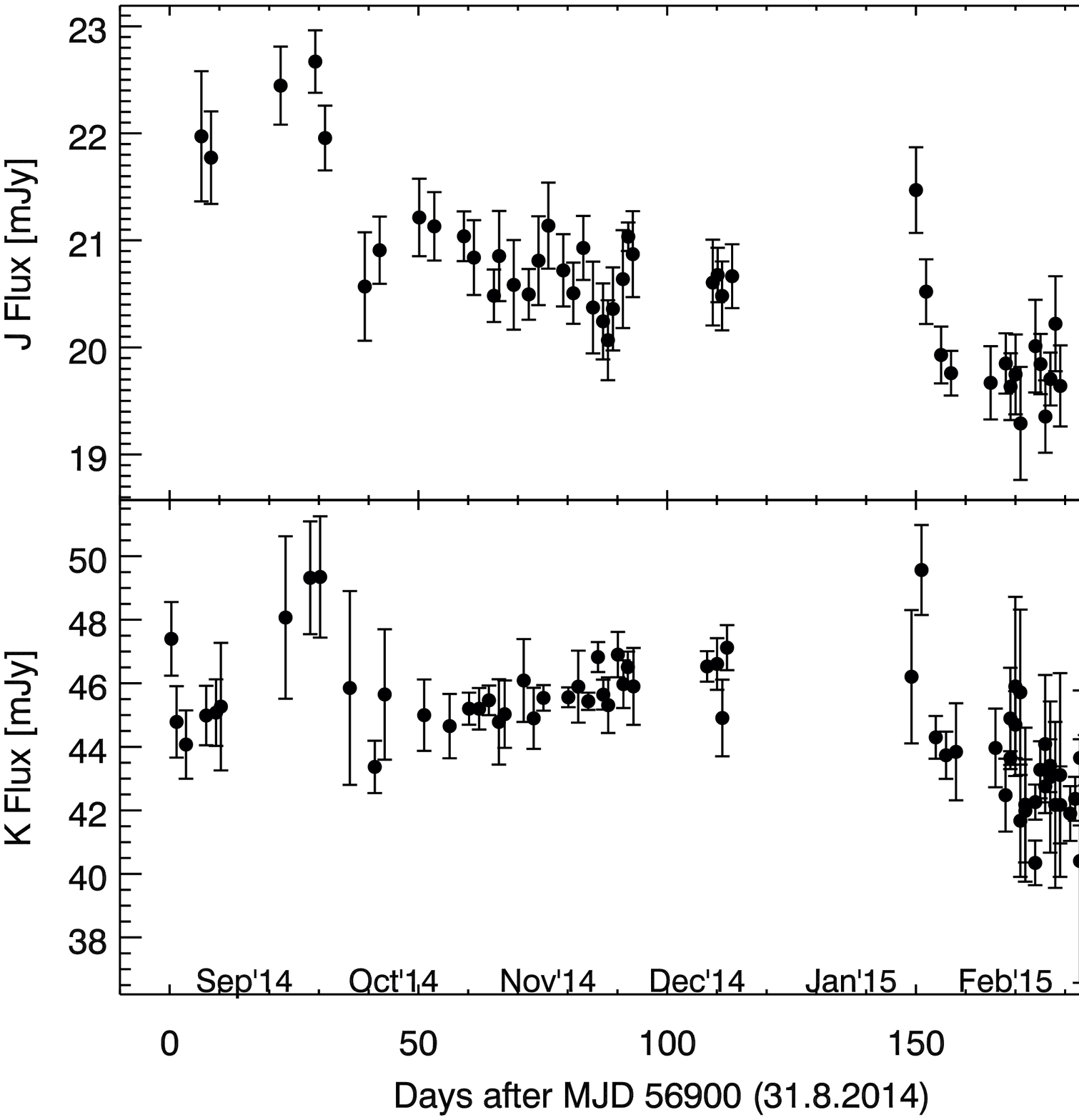}
  \caption{$7\farcs5$ aperture light curves in $J$ and $K$ band. }
  \label{fig:lightcurves_IR}
\end{figure}

The infrared $J$ and $K$ light curves are shown in
Fig. \ref{fig:lightcurves_IR}. In the period from September until
December 2014 one observes a slight decline in the $J$ band similar to
the slope in the optical, but different from the optical bands
(Fig. \ref{fig:lc_2014}), there is no clear rise in the January to
March data. The light curve minimum is reached in February 2015.  The
$K$-band flux also has its lowest flux values in February 2015, but in
contrast to all other bands there is no adequate decline in late 2014,
instead the flux is  rising.
The variations in both bands show a significant delay. The trough of
the large variable feature seen in the optical at November 2014 is now
seen at the end of February 2015.

\label{ssec:dust-rm}

\begin{figure}
  \centering
  \includegraphics[width=\columnwidth,clip=true]{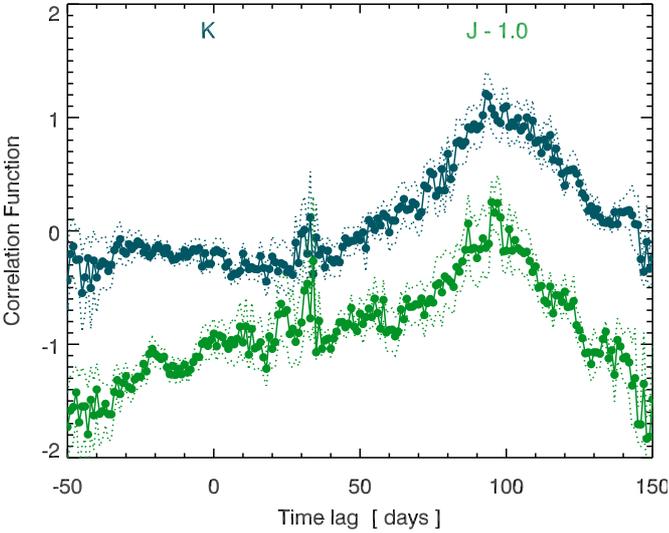}
       
  \caption{Discrete correlation functions between $B$ and IR
    data. The correlation of $B$ with $K$ band is shown in blue; $B$ to
    $J$ band correlation is shifted down by -1 and shown in
    green. Dotted lines represent the 1$\sigma$ uncertainties. }

  \label{fig:dcf_IR}
\end{figure}

\begin{figure} 
  \centering
  \includegraphics[width=\columnwidth,clip=true]{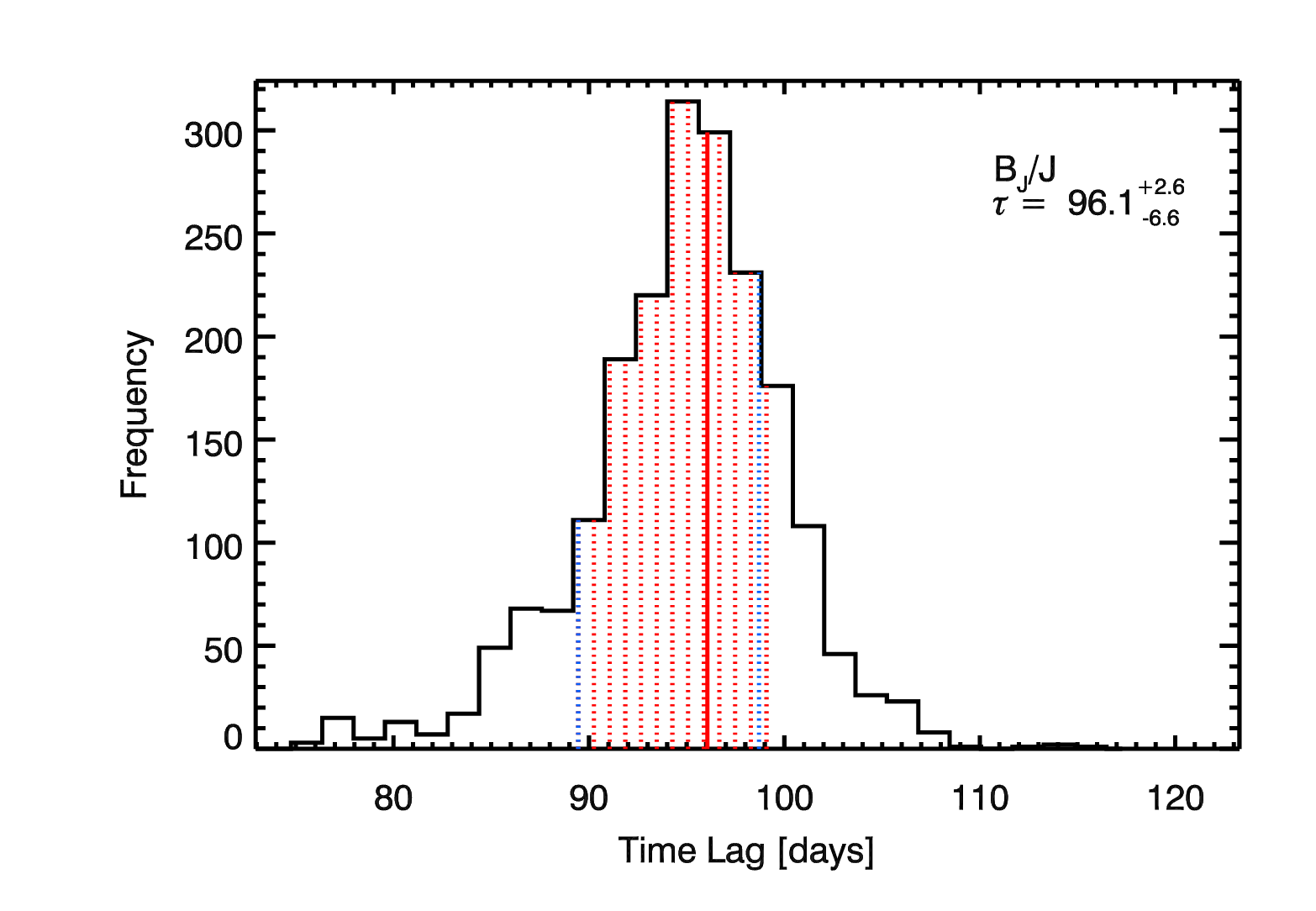}
  \includegraphics[width=\columnwidth,clip=true]{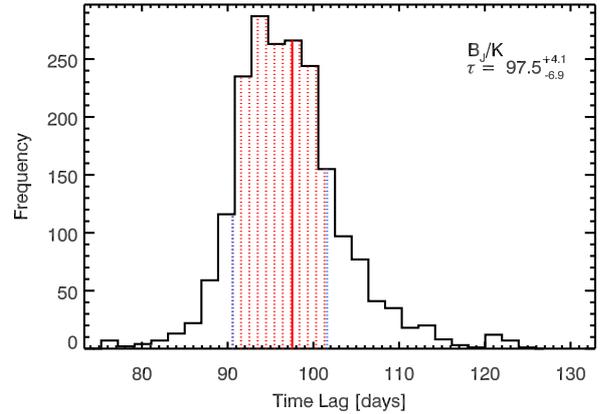}
  \caption{ $B-J$ and $B-K$ DCF statistics
    obtained from 2000 FR/RSS runs. The colored lines indicate median,
    confidence range, and uncertainty as in Fig.~\ref{fig:dcf_BHa}.
    We note the pronounced symmetry of the $B-J$ and $B-K$ distributions
    compared to the asymmetry of $B - H\alpha$ in
    Fig.~\ref{fig:dcf_BHa}.  }
  \label{fig:dcf_BJ}
\end{figure}

We perform the DCF analysis on the IR data. The result is shown in
Fig. \ref{fig:dcf_IR}. In analogy to the analysis in the optical we
apply the FR/RSS method to estimate the errors in the time delays. The
results, corrected for redshift dependent time dilation, are listed in
Tab. \ref{tab:tau_fluxes}.\footnote{{ There are small peaks at the lag
    of about 33 days in both $B-K$ and $B-J$ correlations
    (Fig. \ref{fig:dcf_IR}).  We think that broad emission lines like
    Paschen are not able to explain these small peaks because the
    line strengths are too small (cf. \citealt{2008ApJS..174..282L}
    for typical Sy-1 galaxies) and any Paschen delay would be much
    larger than 33 days, when compared/similar to H$\alpha$ lags.
    Instead, we suspect that the small peaks are caused by the two
    (probably too high) noisy data points in the $J$ and $K$ light
    curves at end of January 2015 (Fig.~\ref{fig:lightcurves_IR})
,    which interfere with the local maximum of the $B$ band light
    curve in December 2014 (Fig.~\ref{fig:lc_2014}).  The small peaks
    in the $B-K$ and $B-J$ correlations essentially disappear, when we
    omit the two $J$ and $K$ data points at end of January 2015 from
    the light curves.}}

In Fig. \ref{fig:dcf_IR}, $B$-band to $J$- and $K$-band correlation
functions are shown, shifted by 1.0 for visibility. This allows us to
see two important aspects of the correlation between optical and IR.
First, the peak of the maximum correlation remains within the FR/RSS
error margins in Tab. \ref{tab:tau_fluxes}.  Second, the
autocorrelation contribution of AD continuum in the IR fluxes becomes
visible at $\tau=0$. This effect is stronger in the $J$ band, which
can be explained by the flux composition in the filters. The $K$ band is
almost completely dominated by the warm dust, while the contribution
of the AD continuum adds a measurable autocorrelation contribution in
the uncorrected $J$ band data. However, the lag of the dust emission
is clearly seen in the peak between 80 and 110 days.  This is further
corroborated by the DCF FR/RSS statistics shown for $B-J$ and $B-K$ in
Fig.~\ref{fig:dcf_BJ}.  We note the strong symmetry in contrast to the
asymmetric $B-H\alpha$ DCF FR/RSS statistics
(Fig.~\ref{fig:dcf_BHa}).\footnote{We have tentatively disentangled
  the contributions of host and AD to the NIR light curves, but this
  requires an uncertain extrapolation of host and AD from the optical
  to the NIR bands. Therefore, we here use the NIR light curves as
  observed.  Nevertheless, in our tests the DCF results for the host
  and AD subtracted NIR light curves become sharper and do not show
  any indication for a substructure as seen for the $B-H\alpha$ DCF
  FR/RSS statistics in Fig.~\ref{fig:dcf_BHa}.  }

\subsection{Dust lag versus luminosity}
\label{ssec:dust_lag_lum}

\begin{figure}
  \centering
  \includegraphics[width=\columnwidth,clip=true]{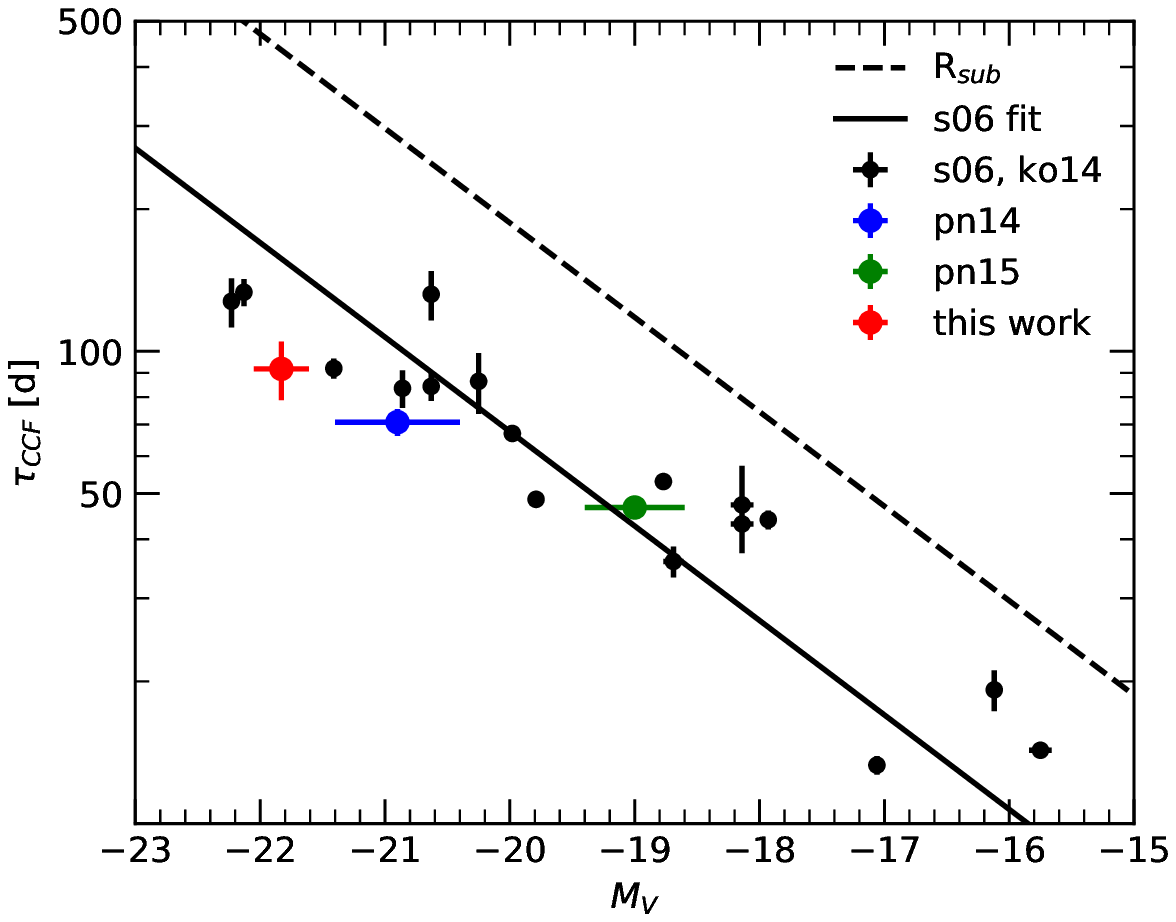}
  \caption{Absolute visual magnitude M$_{\rm V}$ and measured
    cross-correlation delay of the hot dust.  The host-subtracted AGN
    luminosity, corrected for galactic foreground extinction, is
    M$_{\rm V}$ (\citealt{2011ApJ...737..103S}).  Black points are
    based on the data of \cite{2006ApJ...639...46S}, as tabulated in
    \cite{2014ApJ...788..159K}, and labeled s06 and ko14. The
    continuous line is the best fit determined by s06. The black
    dashed line corresponds to the light travel time to the expected
    sublimation radius determined by \cite{2007A&A...476..713K}. Blue
    pn14 data is WVPS\,48 taken from \cite{2014A&A...561L...8P} and
    green (pn15) is PGC\,50427 measured in
    \cite{2015A&A...576A..73P}. Our new measurement for \C is shown in
    red. }
  \label{fig:koshida_comp}
\end{figure}

The dust lag determined with our $B-K$ cross-correlation analysis
places \C on the lower side of the $\tau-M_V$ distribution when
compared with a larger sample by \cite{2006ApJ...639...46S}. This is
depicted in Fig. \ref{fig:koshida_comp}, based on data by
\cite{2014ApJ...788..159K}. \cite{2007A&A...476..713K} found that the
measured average dust lag (solid black line) was a factor of 3 lower
than the expected sublimation radius $R_{\mathrm{sub}}$
(\citealt{1987ApJ...320..537B}; dashed line), which is defined for a
sublimation temperature of 1500\,K.  The solution to this problem
might be found in the geometric shape of the torus as proposed by
\cite{2010ApJ...724L.183K,2011ApJ...737..105K}.  We consider this
further in Sect.~\ref{sec:bowl_geometries}.

  Notably, the four most luminous sources ($M_V < -21\,mag$) lie
  clearly below the average (solid line in
  Fig.~\ref{fig:koshida_comp}).  For 3C\,120 at the given
  $M_V = -21.8\,mag$ Fig.~\ref{fig:koshida_comp} suggests that the
  sublimation radius $R_{sub}$ corresponds to a lag $\tau_{CCF}$ of
  about 434\,days (rest frame, i.e., 450\,days observed frame), hence
  even a factor of 5 larger than measured.  One reason could be that
  after the strong brightening of 3C\,120 in 2013, the dust torus did
  not expand within the one year until 2014 when our observations were
  performed.  Then it would have still a rather small size, controlled
  by the past 1.4 mag lower luminosity state.  For instance, at
  $M_V = -20.4\,mag$ the corresponding lag of $R_{sub}$ would be about
  300 days, consistent with the average line (dashed line in
  Fig.~\ref{fig:koshida_comp}).  A slow receding of the dust torus on
  a timescale of several years also appears consistent with the
  observations of NGC\,4151 by \cite{2013ApJ...775L..36K}.
  {\cite{2004ASPC..311..169L} discussed the timescales of the dust
    sublimation and dust reformation followed by the time variation
    of the illumination, and he noted the possibility of hysteresis
    effects.}  We return to that point in
  Sect.~\ref{ssec:km_bowl}.

\section{Comparison with bowl-shaped geometries}
\label{sec:bowl_geometries}

The $B$-\Ha cross correlation shows a broad maximum plateau ranging
from 55 to 95 days (Fig.~\ref{fig:dcf}), while the $B$-NIR cross
correlations are more peaked ranging from 80 to 110 days
(Fig.~\ref{fig:dcf_IR}).  Thus, at first glance, the BLR lies closer
to the AD than the dust torus, in agreement with the AGN unified
model; the BLR may overlap with or transition into the dust torus.
However, the striking feature is that the dust echo is sharper than
the \Ha echo.  This is clearly not consistent with a bagel-shaped dust
torus, and therefore special geometries have to be considered.  We
compare our data with the bowl-shaped geometry as proposed by
\cite{2010ApJ...724L.183K, 2011ApJ...737..105K}
(Sect.~\ref{ssec:km_bowl}) and the variant bowl model proposed by
\cite{2012MNRAS.426.3086G} (Sect.~\ref{ssec:gkr_bowl}).  A
comprehensive modeling is beyond the scope of this paper, but we
compare the data with a few model cases in this work.
  
Because the inclination for the disk equatorial plane of 3C\,120 is
small, we first consider the models with a face-on view,
i.e., $i=0^{\circ}$.  Thereafter we discuss the effects of inclination,
adopting an inclination of $i=16^{\circ}$, based on the radio jet
orientation found by \cite{2012ApJ...752...92A}.  We note that for a
moderately inclined bowl, the face-on models roughly hold for the left
and right part of the inclined bowl, and that essentially only the
tilted top and bottom parts of the bowl contribute to altering
the transfer function.

In addition, we incorporate gas clouds and the H$\alpha$ BLR inside
the bowl with an estimated covering half-angle of 40.75$^\circ$. We
derive this by assuming a dust covering fraction of $\sim 10\%$
(average between \cite{2011MNRAS.414..218L} and
\cite{2011ApJ...737L..36M}) with a torus half-opening angle of
45$^\circ$, which is consistent with the Sy1/2 number counts
(\citealt{1992ApJ...393...90H}).

A common feature of the bowl geometry is that the height of the BLR
above the equatorial plane increases strongly with equatorial radius
$R_x$.  In Sect.~\ref{ssec:scale_height} we consider how far this fits
the $H/R_x$ estimates derived, if one applies the line profile
analysis proposed by
\cite{2011Natur.470..366K,2013A&A...549A.100K,2013A&A...558A..26K}.

\subsection{Kawaguchi-Mori-Bowl}
\label{ssec:km_bowl}

\begin{figure}
  \centering 
  \includegraphics[width=\columnwidth,clip=true]{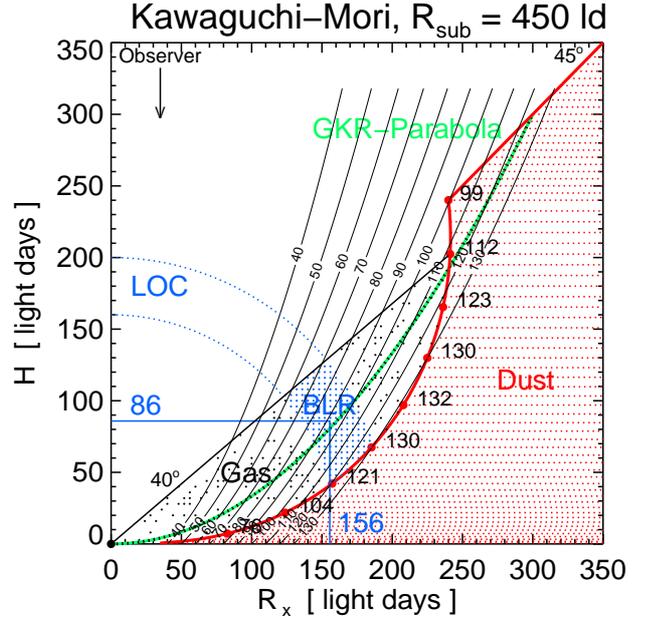}
  \caption{Geometric properties of the dust torus model by Kawaguchi
    \& Mori when seen face-on.  In three dimensions, the model is
    rotationally symmetric around the $y-$axis and mirrored on the
    $x-$axis.  The BH is located at the origin (0,0).  The
    adopted sublimation radius $R_{sub}$ is 450\,ld in polar
    direction.  The thick red line indicates the border of the dust, and
    the volume filled by dust is shaded in red.  The dust distribution
    is cut at a half-opening angle of 45$^\circ$.  For later
    comparison, the green line (labeled GKR-Parabola) indicates the
    slightly different bowl shape as given by
    \cite{2012MNRAS.426.3086G}.  The black curves show the isodelay
    surfaces with numbers giving the delay in days.  The large black
    numbers along the bowl rim list the delay at that place.  The
    black solid line at angle 40$^\circ$ to the equatorial plane
    indicates the potential distribution of the gas clouds (small
    black dots).  The blue hatched area indicates \Ha BLR that we
    see, adopting the LOC model of \cite{1995ApJ...455L.119B}.  The
    thick blue lines indicate the average height (86 ld) and equatorial
    radius (156 ld) of the BLR.  }
  \label{fig:km_bowl_450}
\end{figure}

\begin{figure}
  \centering 
  \includegraphics[width=\columnwidth,clip=true]{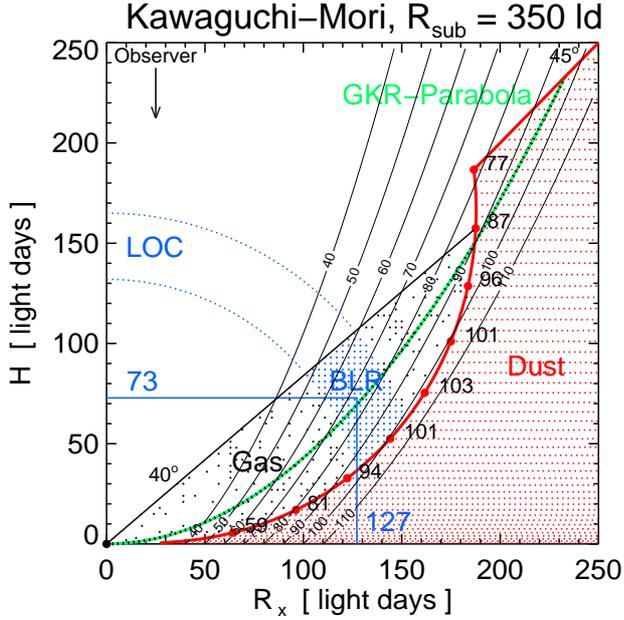}
  \caption{Similar as Fig.~\ref{fig:km_bowl_450}, but for a smaller
    sublimation radius $R_{sub}$ of 350 ld in polar direction.  Also,
    the radius of the LOC has been slightly diminished to 150 ld, and
    the covering half-angle of the gas clouds has been enlarged to
    45$^\circ$.  This places the BLR within the isodelay surfaces
    between 40 and 100 days as indicated by the $B-H\alpha$ cross
    correlation (Fig.~\ref{fig:dcf}).  }
  \label{fig:km_bowl_350}
\end{figure}

We use the dust torus model by \cite{2010ApJ...724L.183K} (henceforth
called KM model), where the rim of the torus is given by the
sublimation radius under the assumption of anisotropic illumination by
the AD (their Eq.~3).  We start adopting an inclination $i=0^\circ$
and sublimation radius $R_{sub}$ in polar direction of 450\,ld
(observed frame), as inferred from Fig.~\ref{fig:koshida_comp}
(Sect.~\ref{ssec:dust_lag_lum}). We adopt the model with
$\theta_{min}=45^\circ$ and $\theta_{max}=89^\circ$.  In addition, we
incorporate gas clouds and the H$\alpha$ BLR inside the bowl with a
tentative covering half-angle of 40$^\circ$.  Guided by the locally
optimized emitting clouds (LOC) model of \cite{1995ApJ...455L.119B},
we place the BLR at a distance of $r_{LOC} \approx 160-200$~ld from
the BH, which corresponds to about 1/2 -- 1/3 of
$R_{sub}$.  We consider this choice as reasonable because it is
derived from average ratios between simultaneous dust and \Ha BLR lag
measurements (see, e.g., \citealt{2014A&A...561L...8P,
  2015A&A...576A..73P}); given a typical ratio lag(\Ha)/lag(H$\beta$)
$\approx$ 1.54 (\citealt{2010ApJ...716..993B}), this choice is also
consistent with the findings of
$\tau_{\rm dust} \approx 4 \times \tau_{H\beta}$ between hot-dust
continuum emission and optical continuum fluctuations
(\citealt{2014ApJ...788..159K}).
         
With regard to the LOC model, however, we mention the following
caveat: For simplicity, the BLR region is placed at the same distance
from the BH for various angles, i.e., not taking into account an
angle-dependent illumination of the BLR clouds by the AD.  For
comparison, an angle-dependent illumination would result in a larger
BLR distance toward the polar direction (compared to the equatorial
direction).  Nevertheless, the BLRs adopted in this work fill only a small
angle range between about 20 and 40 degrees from the equatorial plane
(indicated by the blue shaded areas in Figures 13--16) and therefore
this simplification does not alter the basic results and conclusions
obtained here.

Figure~\ref{fig:km_bowl_450} shows the bowl-shaped dust torus and the
BLR clouds located inside the bowl.  The main features are as follows:
\begin{itemize}
\item [$\bullet$] A large portion of the bowl rim coincides with the
  isodelay surfaces of 120--130 days.\footnote{As bowl rim we consider
    only the curved part of the bowl because it appears reasonable to
    assume that the dust emission outside the bowl at half-opening
    angle 45$^\circ$does not contribute significantly to the $J-$ and
    $K$-band flux variation.  }
            
\item [$\bullet$] If all the dust along the bowl rim contributes to
  the measured varying dust emission, then the expected lag lies
  between 99 and 132 days, hence about 20 days larger than the
  observed range of 80--110 days (Fig.~\ref{fig:dcf_IR}).

\item [$\bullet$] The BLR matches a broad range of isodelay surfaces.
  However, the lags between 60 and 130 days are larger than the
  observed range of 55-95 days (Fig.~\ref{fig:dcf}).  To match the
  observed range, the BLR has to be located closer to the BH, for example, at
  a LOC radius of about $r_{LOC} \approx 50$ ld, but this leads to an
  exceptionally small $r_{LOC} / R_{sub} \approx 0.1$.
\end{itemize}
In conclusion, the model with $R_{sub}$ = 450 ld provides
qualitatively a good approach but fails in detail too much to agree
with the data.  Therefore, we consider a similar model but with
smaller $R_{sub}$ = 350 ld and smaller $r_{LOC} \approx 150$ ld, as
shown in Fig.~\ref{fig:km_bowl_350}.  The main features are as follows:
\begin{itemize}
\item [$\bullet$] A large portion of the bowl rim coincides with the
  isodelay surfaces of 80--110 days, consistent with the data
  (Fig.~\ref{fig:dcf_IR}).

\item [$\bullet$] Likewise, the BLR matches the range of isodelay
  surfaces between 50 and 100 days, consistent with the data
  (Fig.~\ref{fig:dcf}).
\end{itemize}
While this model choice appears satisfying, we did not yet account for
the inclination of $i=16^{\circ}$ which broadens the BLR and dust
transfer functions.  \cite{2011ApJ...737..105K} presented dust
transfer functions for a range of inclinations ($\theta_{obs}$, see
their Fig.~5).  Already at inclinations $10^{\circ}<i<20^{\circ}$ the
dust transfer function broadens, with a centroid at about 0.3 -- 0.4
$R_{sub}$, which translates to a lag of 105 -- 140 days for 3C\,120,
hence larger than what our data measure.  In addition, the dust
transfer function becomes asymmetric with a pronounced maximum at the
long delay end and a tail toward the short delay end.  Our
observational data, however, show very symmetric $B/J$ and $B/K$
cross correlations (Figs.~\ref{fig:dcf_IR},~\ref{fig:dcf_BJ}),
suggesting rather a symmetric dust transfer function.

\begin{figure}
  \centering 
  \includegraphics[width=\columnwidth,clip=true]{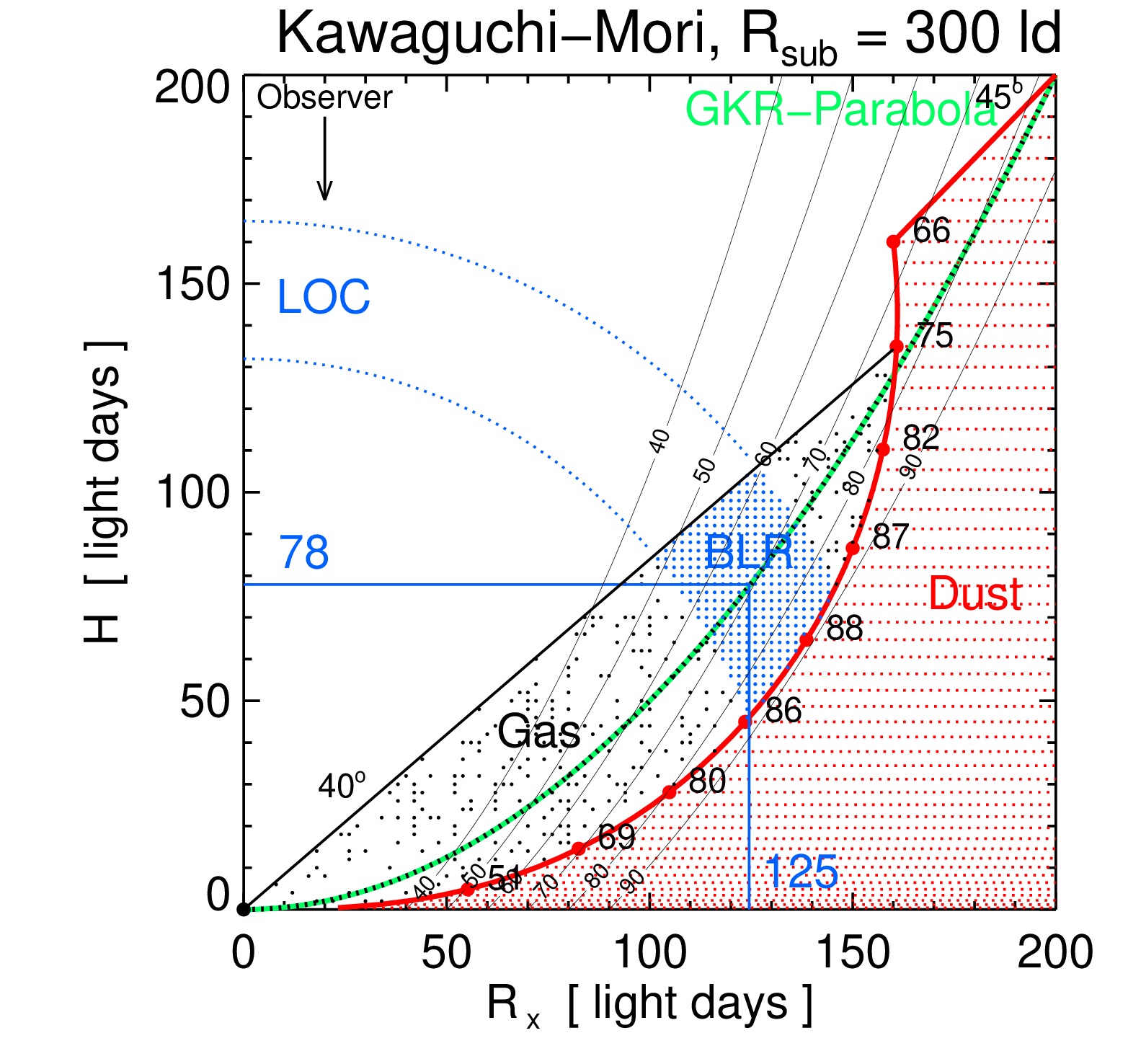}
  \caption{Similar as Fig.~\ref{fig:km_bowl_350}, but for a smaller sublimation 
    radius $R_{sub}$ of 300 ld in polar direction.}
  \label{fig:km_bowl_300}
\end{figure}

Fig.~\ref{fig:km_bowl_300} shows the Kawaguchi-Mori model with even
smaller $R_{sub}$ = 300 ld, which may bring the observed and predicted
lag ranges at $10^{\circ}<i<20^{\circ}$ to an agreement.  However, it
does not solve the discrepancy of the observed symmetric cross
correlation of the light curves and the predicted asymmetric transfer
function.  Therefore, we suggest that other bowl profiles are required
or that the varying hot dust emission does not arise from all along
the entire bowl rim, rather only from a small portion of the rim.  We
discuss further details in Sect.~\ref{sec:discussion}.

\subsection{Goad-Korista-Ruff Model}
\label{ssec:gkr_bowl}
e
  grid in Fig.~\ref{fig:koll_fig16} will soon become publicly
  available (Koll
\cite{2012MNRAS.426.3086G} modeled the BLR, assuming that it is
confined by a bowl-shaped torus geometry and suggesting that the BLR
clouds lie above and next to the bowl surface.  In this model
(henceforth called GKR model), the BLR clouds may shield part of the
dust rim from the AD radiation, reducing the hot dust covering
fraction and bringing it closer to that of $\sim 10\%$ deduced by
\cite{2011MNRAS.414..218L} and \cite{2011ApJ...737L..36M}.  Then the
hot dust emission would arise essentially from the edge, i.e., the
top rim of the bowl.  We use this model with the proposed
parabola-shaped bowl rim, whereby the bowl goes up to a half-opening
angle of 45$^\circ$.  We note that this bowl shape is less convex than
the Kawaguchi-Mori-Bowl (cf. the green line labeled GKR--Parabola in
Figures~\ref{fig:km_bowl_450}--\ref{fig:km_bowl_300}).  This less
pronounced convexity may be interpreted as physical manifestation of
the dust shielding by the gas clouds.  We assume further that the gas
clouds (and the BLR) are located within a half-covering angle $\theta$
of 40$^\circ$, so that the edge lies at
$40^\circ < \theta < 45^\circ$.  We started with an inclination
$i = 0^\circ$. We adjusted the bowl size{, so} that the edge
essentially coincides with the isodelay contour of 100 days.

\begin{figure}
  \centering 
  \includegraphics[width=\columnwidth,clip=true]{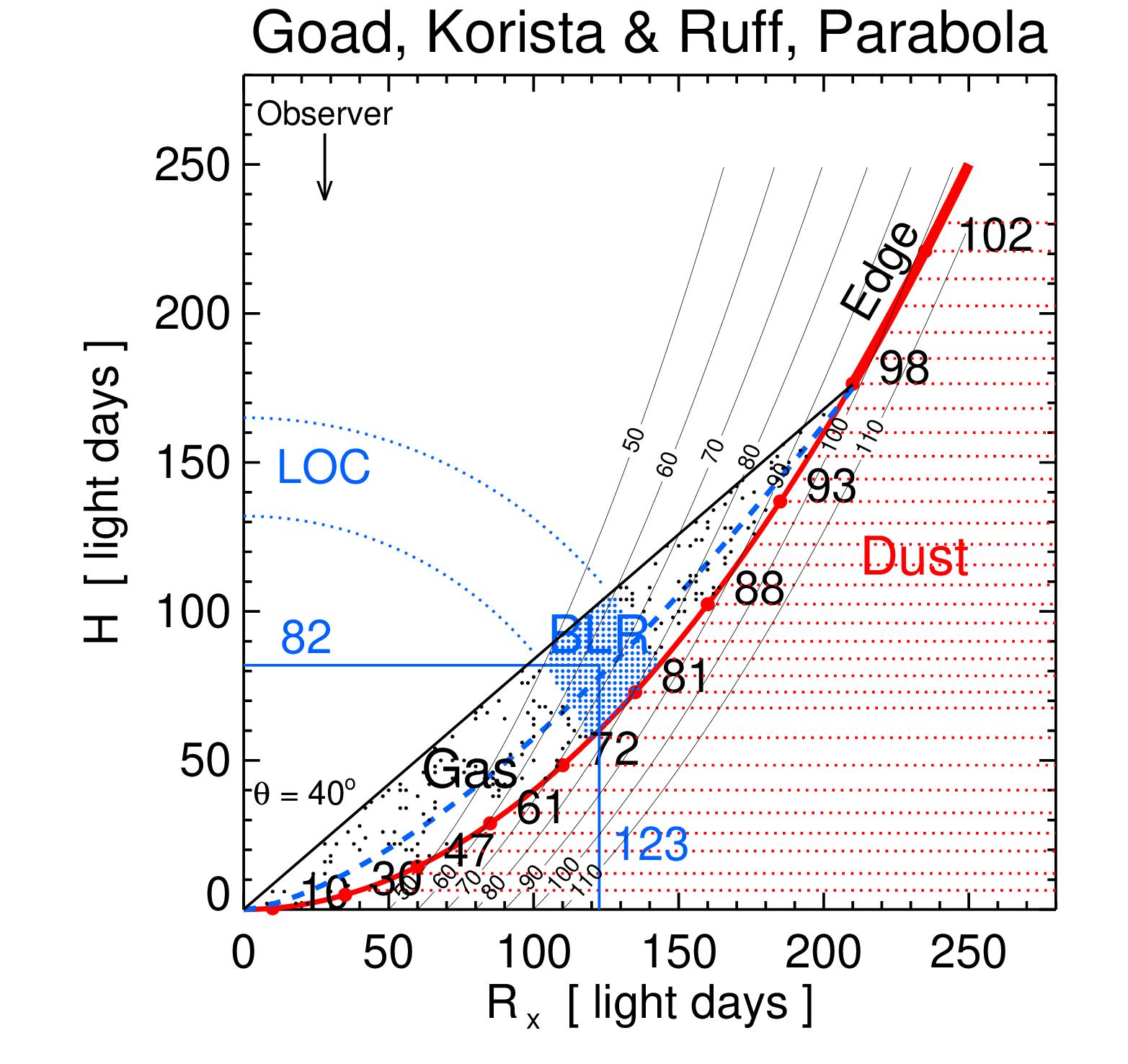}
  \caption{Bowl-shaped torus model with parabolic dust rim, as
    proposed by Goad, Korista \& Ruff (2012).  The lines, colors, and
    labels are similar as for Fig.~\ref{fig:km_bowl_450}.  The very
    thick red line labeled ``Edge'' indicates the dust rim above
    $\theta=40^\circ$.  The thick blue long-dashed line indicates a
    potential BLR gas distribution close to the bowl rim; Goad et
    al. also assumed a LOC controlled BLR gas distribution in their
    calculation of the BLR transfer function.  }
  \label{fig:gkr_bowl_250}
\end{figure}

\begin{figure} 
  \centering
  \includegraphics[width=\columnwidth,clip=true]{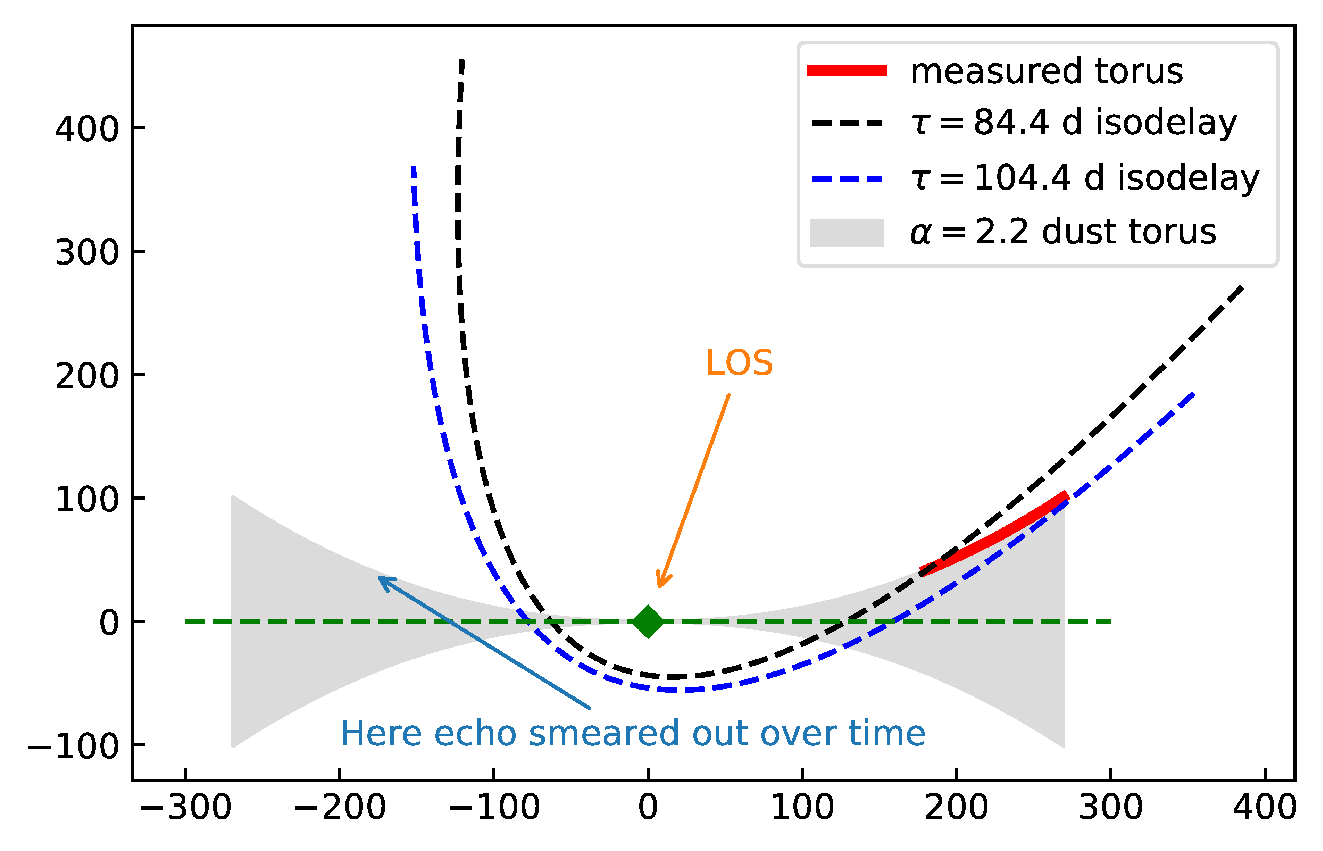}
  \caption{Scheme of the low-$\tau$ peak response and the smeared echo
    for a bowl rim observed at inclination $i\approx 16^\circ$.  The
    dust torus is shaded in gray.  The bowl rim has been adopted as
 a    ``parabola'' with exponent $\alpha = 2.2$, slightly more convex
    than in Fig.~\ref{fig:gkr_bowl_250}.  The long dashed lines indicate
    the isodelay contours.  That part of the rim for which the lag
    matches the data is defined in red; it leads to a sharp low-$\tau$
    peak response.  On the other hand, the entire part of the rim on
    the left-hand side (marked by the blue arrow) leads to a broad
    response that is more delayed and smeared out over time.  }
  \label{fig:isodelay_example}
\end{figure}

\begin{figure}
  \centering
  \includegraphics[height=\columnwidth, angle =-90]{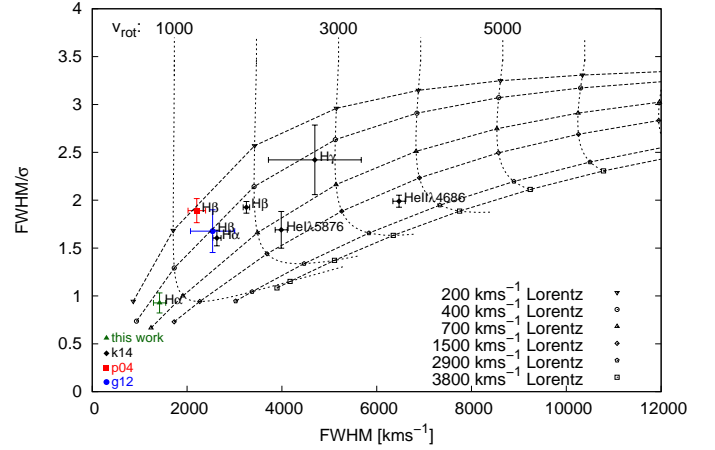}
  \caption{Ratio of FWHM/$\sigma$ with our observed value and
    measurements of previous studies of \C and other lines; p04:
    \cite{2004ApJ...613..682P}; g12: \cite{2012ApJ...755...60G}; k14:
    \cite{2014A&A...566A.106K}. Dashed lines indicate theoretical ratios
    of rotationally broadened Lorentzian profiles.}
  \label{fig:koll_fig16}
\end{figure}

Figure~\ref{fig:gkr_bowl_250} shows the bowl-shaped dust torus at
inclination $i=0^\circ$ and the BLR clouds located inside the bowl.
The main features of the dust and BLR echoes are as follows:
\begin{itemize}
\item [$\bullet$] Dust: Below the edge, the lag decreases steadily for
  the inner parts of the bowl rim.  The rim crosses many isodelay
  surfaces.  If all the dust along the bowl rim contributes to the
  measured varying dust emission, then the expected dust lag lies
  between about 61 and 102 days.  We note that for an inclined bowl
  (say at $i=16^\circ$; see \citealt{2012ApJ...752...92A}) the
  transfer function becomes broader than at $i=0^\circ$.  Then the
  expected dust lags disagrees with the sharp NIR echo observed.  This
  argues for a scenario in which the gas clouds shield the inner bowl
  rim below the edge and in which the edge provides the bulk of the
  varying hot dust emission, as proposed by Goad et al.
            
\item [$\bullet$] BLR: The BLR (blue shaded area, estimated by the LOC
  principle) matches the isodelay contours at $i=0^\circ$, which
  appears consistent with the observed range of 55-95 days (see
  Fig.~\ref{fig:dcf}).  If the bowl is inclined at $i=16^\circ$, this
  range of lags holds for the left and right part of the inclined
  bowl, while the tilted top and bottom smears out the lag range
  as illustrated in Fig.~\ref{fig:isodelay_example}.  We now consider
  this effect further.

  In their Fig.~1, Goad et al. assumed a gas distribution rather close
  to the bowl rim, similar to that marked by the long dashed blue line
  in Fig.~\ref{fig:gkr_bowl_250}. While at $i=0^\circ$ the expected
  BLR lag range appears then very narrow, it broadens with increasing
  inclination.

  \cite{2012MNRAS.426.3086G} show calculated BLR transfer functions
  (response functions $\Psi(\tau)$) for a grid of parameters and
  inclinations in their Fig.~5; their BLR calculations took the LOC
  principle into account.  For the parabola (exponent $\alpha$ = 2)
  they find: At $i=10^\circ$: $\Psi(\tau)$ is symmetric with clear
  $\tau$ peak and extended wings.  At $i=30^\circ$: $\Psi(\tau)$ shows
  a strong asymmetry, with a pronounced peak at low $\tau$ (with a
  value of about 20\% of the median of the $\tau$ range).  By
  eye-balled linear interpolation, we estimate that at $i=16^\circ$
  the asymmetry shows up as well with a low-$\tau$ peak at about 80\%
  of the median of the $\tau$ range, exactly as seen in the data
  (58/72, see Fig.~\ref{fig:dcf_BJ}).  Thus, the GKR BLR model appears
  nicely consistent with our 3C\,120 data.
\end{itemize}
To summarize, the GKR model has the interesting property that -- by
shielding the dust rim by the gas clouds -- places the hot dust
emission to the edge, i.e., a spatially confined region, allowing for a
rather sharp and symmetric dust transfer function at small
inclinations as well; this is consistent with the 3C\,120 observations.  At the same
time, the GKR model predicts a broad transfer function for the BLR
with a low-$\tau$ peak, which is again consistent with the 3C\,120
observations.

\subsection{FWHM/sigma and the scale height of the BLR}
\label{ssec:scale_height}
In a bowl geometry the scale height $H$ of the BLR above the
equatorial plane increases strongly with equatorial radius $R_x$.
We consider how far this fits the $H/R_x$ estimates derived from
FWHM/sigma\footnote{sigma is the line dispersion}, if one applies the
line profile analysis proposed by
\cite{2011Natur.470..366K,2013A&A...549A.100K,2013A&A...558A..26K}.
In particular, this analysis separates rotational and turbulent
velocities ($v_{\rm rot}$ and $v_{\rm turb}$, resp.), allowing us to
derive an improved BH mass using a proper $v_{\rm rot}$.  We
begin with a few introductory explanations.

While the BLR kinematics is widely believed to be gravitationally
dominated by the BH (e.g.,
\citealt{1988ApJ...325..114G,1999ApJ...521L..95P}), a thin disk-like
geometry with a pure Keplerian velocity profile is not able to explain
most of the observed AGN spectra.  \cite{1978PNAS...75..540O} already
noted that some vertical extension is required for the BLR and
consequently some significant vertical motion to maintain its
thickness. Fitting the statistics of line widths, he found a typical
rotational velocity of 5000 km/s and a turbulent velocity of 2000 km/s
for the BLR gas.

It is well known that the FWHM/sigma line ratio of the broad emission
lines in the best reverberation studied Seyfert-1 galaxies and quasars
shows a large dispersion (\citealt{2004ApJ...613..682P}).
\cite{2006A&A...456...75C} investigated the FWHM/sigma line ratio of
that sample.  These authors showed that this ratio is anticorrelated with
the Eddington ratio and with line widths; broader emission lines
(Population 1) tend to have relatively flat-topped profiles and
narrower lines (Population 2) have more extended wings (see also
\citealt{2000ApJ...536L...5S, 2004A&A...426..797C}).  Collin et
al. tentatively suggested that the difference between the two
populations is due to the relative strength of a disk-wind component,
which is stronger in Population 1. These authors discuss the possible role of
self-gravity as the physical driver controlling the strength of the
disk wind. They find a stronger wind is expected for larger Eddington
ratios, which is consistent with the smaller FWHM/sigma line ratios
found for Population 1 objects.\footnote{A caveat should be mentioned:
  a small line width leads to a small BH mass; this
  subsequently leads -- at a given luminosity -- to a larger Eddington
  ratio (\.M/M$_{\rm BH}$).  Thus a degenerated anticorrelation
  between line width and Eddington ratio is inherently predicted.}

In a different approach,
\cite{2011Natur.470..366K,2013A&A...549A.100K} analyzed broad emission
line profiles of C\ion{IV}, helium, and Balmer lines of the above
sample presented by \cite{2004ApJ...613..682P}.  They considered
emission line profiles (Gaussian, Lorentzian, logarithmic,
and exponential) resulting from various kinematical and dynamical models.
{For instance,} turbulent motions result in Lorentzian profiles,
inflow or outflow motions lead to logarithmic profiles and electron
scattering to exponential profiles; all three profiles yield
FWHM/sigma~$\approx$~1 and therefore they are further commonly
denoted as turbulence.  Assuming that these motions are superimposed
by the Keplerian motion of the gas clouds, Kollatschny \& Zetzl
computed the rotational convolution of Lorentzian profiles and derived
the FWHM and sigma values of these newly generated convolved profiles.
Finally, they compared FWHM/sigma and FWHM for the data and model
profiles.  This way, they used the FWHM/sigma line ratio versus FWHM
diagram for separating rotational and turbulent velocities. Their
studies so far indicate that inclination plays a minor role and is
limited to a few outlier sources, similar to what
\cite{2006A&A...456...75C} found in another
study.\footnote{\cite{2006A&A...456...75C} considered
  FWHM/sigma versus sigma (see their Fig.~3), which is a 1/x versus
  x dependency (x=sigma) and as such somewhat difficult to
  interpret.  In contrast, \cite{2011Natur.470..366K} analyzed
  FWHM/sigma versus FWHM, which evidently reveals a clear and smooth
  deviation from a linear x vs. x trend (x=FWHM) and finally
  puts the suggested existence of two distinctly separated populations
  into question.  One may speculate that the sources with relatively
  large turbulent velocities by Kollatschny \& Zetzl are exactly those
  Population 1 sources by Collin et al., but this is beyond the scope
  of this paper.  }

\subsubsection{Disentangling rotational and turbulent velocities} 
\label{ssec:scale_height_1}

Fig.~\ref{fig:koll_fig16} shows the FWHM/sigma versus FWHM
measurements of \C from previous campaigns (see Fig.~22 in
\citealt{2014A&A...566A.106K}) together with the \Ha result of our new
epoch, determined through the spectral decomposition in
Sect. \ref{ssec:fast}. The dotted and dashed lines represent the grid
of computed theoretical rotational and turbulent Lorentzian profiles
following \cite{2011Natur.470..366K,2013A&A...549A.100K}.\footnote{Thatschny et al., 2018, in prep.). A preliminary
  version of the grid can be found at
  http://www.astro.physik.uni-goettingen.de/$\,$\~{}zetzl/blrvelo/}

From this diagram we determined the rotational and turbulent
velocities to be
\begin{equation}
  \label{eq:vrot}
  v_{\rm rot}= 770^{+118}_{-158}\,\mathrm{km}\,\mathrm{s}^{-1},\\
  v_{\rm turb}= 574^{+191}_{-166}\,\mathrm{km}\,\mathrm{s}^{-1}
.\end{equation}
Compared to the \Ha value in \cite{2014A&A...566A.106K}, the
rotational component has decreased from
$1526^{+60}_{-66}$\,km\,s$^{-1}$ (i.e., by a factor 2), while the
turbulent component remains unchanged within the errors or has
marginally increased from $548^{+108}_{-101}$\,km\,s$^{-1}$
(Tab.~\ref{tab:bh-masses}).

If $v_{\rm rot}$ manifests the Keplerian motion of the clouds, the
decrease of $v_{\rm rot}$ is qualitatively consistent with the
increase of the \Ha BLR radius: if the BLR lies in the equatorial
plane, the radius $R$ would increase by a factor 2.4 from 28.5~ld to
68.9~ld (Tab.~\ref{tab:bh-masses}). For a BLR in the equatorial plane,
however, the virial mass
\begin{equation}
  \label{eq:vp}
  M_{vir} = R \cdot v_{\rm rot}^2 / G
\end{equation}
would then decrease by a factor $2^{\rm 2}/2.4 = 1.67$, which is
quantitatively incompatible with the obvious requirement of
$M_{\rm vir}$ being constant. The value $M_{\rm vir}$ times a
geometric scaling factor $f$ is equal to the BH mass M$_{\rm BH}$.
This suggests that a special geometry is required to resolve the
discrepancy.  Such a geometry could be a bowl shape with
foreshortening effect.  We discuss that topic further in
Sect.~\ref{ssec:discussion_back_hole_mass}.

\subsubsection{BLR scale height and radius}
\label{ssec:scale_height_2}

In theoretical studies of ADs around a compact central
mass, \cite{1981ARA&A..19..137P} found that the disk scale height $H$
at equatorial radius $R_x$ is given by
\begin{equation}
  \label{eq:height}
  H = (c_{\rm s} / v_{\rm rot}) \cdot R_x 
\end{equation}
with the sound speed $c_{\rm s}$ and the Keplerian velocity
$v_{\rm rot}$ (see his Eq.~3.16).  For an ensemble of particles,
$c_{\rm s}$ and $v_{\rm rot}$ denote the average of the particle
distributions.  Using this relation, \cite{2013A&A...549A.100K} made
the further assumptions that $v_{\rm turb} < c_{\rm s} $,
$v_{\rm turb} = \alpha' c_{\rm s}$, where $\alpha'$ is close to the
usual viscosity parameter $\alpha \le 1$
(e.g., \citealt{1973A&A....24..337S}).  If the BLR gas clouds are
distributed in such an AD-like configuration, then we can
compute the scale height of the BLR as
\begin{equation}
  \label{eq:ratio}
  H = (1 / \alpha)  \cdot (v_{\rm turb} / v_{\rm rot}) \cdot R_x
.\end{equation}
For Keplerian orbits (with $v_{\rm rot}^2 \cdot R_x$ =
constant) this yields
\begin{equation}
  \label{eq:prop}
  H \propto  (1 / \alpha) \cdot v_{\rm turb} \cdot R_x^{\rm 1.5} 
.\end{equation}
This is a flared disk in which the height increases over-proportionally
with radius.\footnote{For $\alpha \ll 1$, say
  $\alpha = 0.1$, $H/R_x$ may become larger than unity, hence the
  assumption of a geometrically thin AD does no longer
  hold.  Therefore, \cite{2013A&A...549A.100K} assumed
  $\alpha' \approx \alpha = 1$, which is a conservative estimate
  providing a lowest possible $H/R_x$.  } Furthermore, the
gravitational potential of the central mass decreases with increasing
$R$, so that -- for constant $\alpha$ and $v_{\rm turb}$ -- the clouds
at large $R_x$ can lift up to larger height above the equatorial
plane.

For our new \Ha data of 3C\,120, the usage of Eq.~\ref{eq:ratio}, with
values and errors from Eq.~\ref{eq:vrot} and assuming the viscosity
parameter $\alpha$ to be unity, yields
\begin{equation}
  \label{eq:ratio2}
  H = 0.75^{+0.5}_{-0.35} \cdot R_x
.\end{equation}
We compare this spectroscopically obtained value with that
geometrically derived for the bowl-shaped BLRs in
Fig.~\ref{fig:gkr_bowl_250}. If the BLR clouds lie at the intersection
of the 70\,d isodelay contour with the bowl rim, then
$H/R_x \approx 50/100 = 0.5$.  For the blue shaded BLR area $H=82$\,ld
and $R_x = 123$\,ld we obtain $H/R_x = 82/123 \approx 0.67$ (see the
thick blue lines with labels).  Given the uncertainties for both the
spectroscopic and geometric method, the $H/R_x$ values agree
surprisingly well.
          
\section{Discussion}
\label{sec:discussion}

The simultaneous BLR and dust reverberation data of 3C\,120 after
strong brightening by a factor 3 allows us to check for luminosity
dependent effects.  One such effect is the geometry of the BLR--torus
system.  We discuss a bowl-shaped geometry in this section.

\subsection{Change of bowl size with varying luminosity}
\label{ssec:discussion_bowl_varying}
          
When the luminosity increases, we expect that the dust torus size
increases owing to sublimation of the innermost dust and vice versa
(receding torus model; \citealt{1991MNRAS.252..586L}). However, the
timescale on which this increase/decrease happens is not yet known.
It is likely that the receding of the torus is slow, lasting longer
than one year.  Indications for a yearlong timescale come from the
observations of NGC\,4151 by \cite{2013ApJ...775L..36K}.  For \C the
ratios of lag(dust) to lag(BLR) and lag(dust) to $M_V$ are small
compared to other AGN.  This suggests that \C is overluminous for the
current bowl size and that the torus rim has not yet receded in accord
with the current high luminosity.

The AD luminosity of a Seyfert AGN changes typically by 10\%--30\%
within a few weeks.  While the bowl size appears to stay essentially
constant or to change slowly over at least several months, the gas
clouds obviously react fast (within days or hours) to a change in
luminosity.  According to the LOC model of \cite{1995ApJ...455L.119B},
the gas clouds do not instantaneously move away or toward the
illumination source, rather only their excitation and ionization
states change, depending on the distance from the illuminating compact
AD.  This change can be instantaneous and those clouds fulfilling the
locally optimized conditions shine as BLR.  Then at a low luminosity
state the BLR consists essentially of clouds in the bottom of the bowl
with small $H/R_x$, but at a high luminosity state the BLR is
preferably located at larger distance, hence at larger $H/R_x$.  In
other words, the BLR swings inside the bowl with the systematic trend
that $H/R_x$ increases with increasing luminosity due to the bowl rim.
It is possible to further speculate that the BLR covering angle
increases with increasing luminosity as well.  We discuss these issues
further in Sect.~\ref{ssec:discussion_r_l}.


\subsection{Black hole mass calculation}
\label{ssec:discussion_back_hole_mass}

We consider potential effects of the bowl-shaped geometry on the
BH mass calculation.

Reverberation based BH masses are commonly computed using
\begin{equation}
  \label{eq:vir_mass}
  M_{\mathrm{BH}}= (f \cdot R \cdot \Delta v^2) / G
,\end{equation}
where $\Delta v$ is a measure for the velocity, $R = c \cdot \tau$
and $f$ is the geometric scaling factor that is still under
debate. From a statistical analysis of 16 AGN,
\cite{2004ApJ...615..645O} determined $f=5.5$, but also other values
$3<f<9$ have been reported; the gravitational redshift measured for
Mrk\,110 indicates $f=7.8$ in this source if the BLR is an inclined
ring at $i=21^\circ$ (\citealt{2003A&A...412L..61K}).  To be
independent of $f$, we consider the virial mass M$_{\rm vir}$
(Eq.~\ref{eq:vp}), i.e., M$_{\rm BH}$ adopting $f=1$.
          
Table~\ref{tab:bh-masses} lists virial masses from previous
reverberation campaigns of \C and our new data.  For intercomparison,
the virial masses are calculated using several parameter sets:
standard velocity estimates (FWHM, $\sigma_v$) and
$v_{\rm rot}$ as determined from the FWHM/sigma analysis
(Sect.~\ref{ssec:scale_height_1}).  Likewise, two versions of the
orbit radius are used: the standard $R= c\cdot \tau$ (middle block,
rows 7--9 in Tab.~\ref{tab:bh-masses}) and the radius
$R_{corr} = c \cdot \tau / f_{fs}$ corrected for the foreshortening
factor $f_{fs}$ of the measured lag $c \cdot \tau$ (last block, rows
10--14 in Tab.~\ref{tab:bh-masses}).  The calculation of $f_{fs}$
requires knowledge of $H/R_x$.  For the correction we tentatively
adopt the $H/R_x \propto v_{\rm turb} / v_{\rm rot}$ values from the
FWHM/sigma analysis (Sect.~\ref{ssec:scale_height_2}).  We further
adopt a face-on view, $i=0$, which roughly holds for the left and
right side on an inclined bowl and therefore allows us to estimate a
mean $f_{fs}$.  Then $f_{fs} = 1 - sin(\theta)$ with
$\theta = atan(H/R_x)$.  For the parameters in the last block the
errors are very uncertain and therefore not listed; nevertheless the
numbers allow us to recognize trends.  Thus, Table~\ref{tab:bh-masses}
lists viral masses obtained for four different methods (rows 7--9,
14).

We begin with considering the previous data, first at low luminosity
for \Hb, then at intermediate luminosity for \Ha and \Hb, and finally
the new data at high luminosity.
\begin{itemize}
\item [ 1) ] Columns 2--4 list the results for \Hb obtained at three
  low luminosity campaigns.  For each method the virial masses agree
  notably well.  Figure~\ref{fig:gkr_bowl_250} indicates that
  at a lag of about 25--38~days the \Hb BLR lies presumably in the
  bottom of the bowl at small $R_x < 50$\,ld where the scale height
  and potential foreshortening effects might play a minor role.  Thus
  the differences in the viral masses between different methods are
  due to the different kinds of velocity measures used.
  \smallskip

\item [ 2) ] Columns 5--6 list results obtained for \Ha and \Hb from
  the 2008/2009 campaign by \cite{2014A&A...566A.106K} at 10--25\%
  larger luminosity (row 1 in Tab.~\ref{tab:bh-masses}).  For each of
  the four methods the virial mass of \Ha agrees notably well with the
  previous estimates obtained with the same method.  For \Hb, however,
  all four methods yield a larger virial mass than the previous
  campaigns; within the uncertainties the virial mass may be
  considered consistent.\footnote{We cannot exclude systematic errors
    in the k14 \Hb data, for example, by the contamination of the iron
    lines at the red flank of the \Hb line.}  \smallskip
              
\item [ 3) ] Column 7 list results obtained for \Ha with the new data.
  The parameters differ significantly by about a factor 2 from
  previous parameters: the lag $\tau$ is larger, and the velocities
  FWHM and $v_{rot}$ are smaller.  Using the standard
  $R = c \cdot \tau$ this leads to smaller virial masses compared to
  previous values, in particular for $v_{rot}$ (row 9).  To solve the
  discrepancy of M$_{\rm vir}$, we suggest that the lags do not
  linearly translate into orbit radius, i.e., distance from the BH. In
  a bowl-shaped geometry, \Ha is emitted at largest distance, hence at
  largest scale height $H$, and thus suffers from the largest
  foreshortening.  When correcting for the foreshortening factor using
  the FWHM/sigma analysis, the virial mass using $R_{corr}$ agrees
  well with that obtained in the previous campaigns (row 14).

  However, for this spectroscopically derived $R_{corr}$ we obtain a
  large $\theta = 37^\circ$ close to adopted maximal half-covering
  angle $\theta = 40^\circ$.  Therefore we consider the two geometric
  alternative $H/R_x$ estimates from Fig.~\ref{fig:gkr_bowl_250}: (1)
  if the BLR lies at the bowl rim, where it cuts the isodelay surface,
  then $H/R_x \approx 70/140 = 0.5$. (2) If we adopt the shaded BLR,
  then $H/R_x \approx 82/123 = 0.67$.  These values are smaller than
  0.75 and lead to a more moderate correction and a slightly smaller
  virial mass: $R_{corr}= 124.6$\,ld,
  M$_{\rm vir} = 14.3 \cdot 10^{6} M_\odot$, and
  $R_{corr}= 154.7$\,ld, M$_{\rm vir} = 17.8 \cdot 10^{6} M_\odot$,
  respectively.  Both values agree well with previous M$_{\rm vir}$
  estimates.
\end{itemize}
To summarize, in a bowl-shaped geometry the calculation of
M$_{\rm vir}$ may be affected by the foreshortening of the observed
BLR lag, which leads to an underestimation of the BLR distance from the
BH.  The foreshortening effect increases with increasing
distance of the BLR from the BH.  For the 3C\,120 data, a
tentative correction of the foreshortening effect by spectroscopic or
geometric $H/R_x$ estimates yields consistent M$_{\rm vir}$ values for
different Balmer lines (\Ha, \Hb) and different AGN luminosities.

\begin{table*}
  \begin{center}
    \hfill{}
    \caption{ Virial masses, i.e. black hole masses adopting $f=1$, in
      units of 10$^6$ M$_\odot$ computed for several methods, as well
      as input parameters derived from several campaigns.  p04:
      \cite{2004ApJ...613..682P}; g12: \cite{2012ApJ...755...60G};
      p12: \cite{2012A&A...545A..84P}; k14:
      \cite{2014A&A...566A.106K}; and this work. The velocity
      parameters of p04, g12 and k14 are from the rms spectra, while
      those of p12 and this work are from a single epoch spectrum.  }
              \label{tab:bh-masses}
              \begin{tabular}{@{}r|l|ccc|cc|c}
                \hline
                (0) & (1) & (2) & (3) & (4) & (5) & (6) & (7) \\
                \hline
              row &  line  (campaign)              & \Hb (p04)          & \Hb (g12)          & \Hb (p12)          & \Ha (k14)          & \Hb (k14)          & \Ha (this work)        \\
                \hline
              1&  $log(\lambda L_{\lambda 5100\AA})$ [erg\,s$^{-1}$]  & $44.01 \pm 0.05 $  & $43.87 \pm 0.05 $  & $43.84 \pm 0.04 $  & $44.12 \pm 0.07$   & $44.12 \pm 0.07$   &$ 44.32 \pm 0.07 $  \\
              2&  $\tau$  [days]                & $38.1 \pm 18.3 $   & $25.6 \pm~~2.4 $   & $23.6 \pm~~1.7 $   & $28.5  \pm 8.8 $   & $27.9  \pm 6.5$    & $68.9 \pm  12.9$   \\
                \hline
              &  line widths [km\,s$^{-1}$]:         &           &       &                        &       &   & \\
              3&  FWHM                          & $2205 \pm 185$ & $2539 \pm 466$  & --     &  $2630 \pm 87$ & $3252 \pm 67$ & $1420 \pm 129$ \\
              4&  $\sigma_v$                    & $1166 \pm~ 50$ &  $1514 \pm 65$ &   1504             &  $1638\pm 105$  & $1689 \pm 68$  & $1532 \pm 201$   \\
              5&    $v_{\mathrm{turb}}$                & $ 243 \pm~~49$ & $ 453 \pm 108$ & --  & $548 \pm 105$  & $507 \pm 72$   & $574 \pm 178 $   \\
              6&    $v_{\mathrm{rot}}$                 & $1291 \pm 110$  & $1480 \pm 287$ & --  &  $1526 \pm 63$ & $1901 \pm 44$  &  $ 770 \pm 138$  \\
                \hline
               &    $R = c \cdot \tau$           &           &       &                &        &          & \\
              7&    $M_{\mathrm{vir}}$  (FWHM, $R$)           &   $36.0 \pm 10.0$ & $32.1 \pm 4.2$ &           -- & $38.3 \pm 10.0 $    &  $57.3 \pm 15.0$         &  $27.0 \pm 5.0$ \\
              8&    $M_{\mathrm{vir}}$  ($\sigma_v$, $R$)     &   $10.1 \pm 5.0$ & $12.2 \pm 1.2$ & $10.4 \pm 4.9$ & $14.9 \pm  5.0 $   &  $15.5 \pm  5.0$         &  $31.4 \pm 9.0$ \\
              9&    $M_{\mathrm{vir}}$  ($v_{\mathrm{rot}}$, $R$)  &  $12.3 \pm 5.0$ & $11.4 \pm 1.1$ & --             & $12.9 \pm 4.7$  & $19.6 \pm 6.2$ &  $6.9 \pm 1.0$   \\
                \hline
              10&    $H/R_x = v_{\mathrm{turb}} / v_{\mathrm{rot}}$  &  0.31  & 0.19  &  --  & 0.36       &  0.27     & 0.75 \\
              11&    $\theta =  atan(H/R_x)$ [ $^\circ$ ]              &  17       & 11      &  --  & 20            & 15          & 37    \\
              12&    $f_{fs} = 1 - sin (\theta)$            &  0.70       &  0.81     &  --  & 0.66            & 0.74          & 0.40    \\
              13&    $R_{corr} = c \cdot \tau / f_{fs}$  [ ld ]      & 54.1        & 31.5      &  --  & 43.1            & 37.7          & 172    \\
              14&    $M_{\mathrm{vir}}$  ($v_{\mathrm{rot}}, R_{corr}$)     &  17.5         & 13.4      & -- &  19.5     & 26.4         &  19.8   \\

                \hline
    \end{tabular}
  \end{center}
\end{table*}

\subsection{Evidence favoring a bowl-shaped BLR \& dust geometry}
\label{ssec:discussion_evidence}

The cross correlations of the respective light curves $B$--\Ha and
$B$--NIR yield that the \Ha echo is broad with a pronounced low-$\tau$
  peak (Fig.~\ref{fig:dcf_BHa})
and the dust echo is symmetric and sharper than the
  BLR echo (Fig.~\ref{fig:dcf_BJ}).
 To make these results consistent, we compared the data
with bowl-shaped BLR--torus models by KM and GKR.  We adopted an
inclination $i = 16^\circ$.  The KM bowl is more convex than the GKR
bowl.

The comparison shows,
if the hot dust {emission} comes from the entire
  rim, then, both the KM and GKR models  yield  no good fit of the lag
  to the data.  The predicted asymmetries in the transfer functions
  and CCFs are not consistent with the symmetry of the data.
This led us to the conclusion that the hot varying
  dust seen in the J and K bands is located only at the edge of the
  bowl.  This could be achieved if the BLR gas clouds shield most of
  the dust rim below a covering angle of about 40$^\circ$ (measured
  from the equatorial plane).
The BLR gas clouds are located in the bowl, along
  and above the rim, but not entirely up to the edge.  Then the
  predicted broad transfer function with low-$\tau$ peak is consistent
  with the cross-correlation function of the data ($B -$\Ha).
 Regarding the precise parameterization of the bowl
  rim, we cannot exclude the KM model by our eyeball comparison of
  the data with a few $R_{sub}$ cases. {We also mention the caveat
    that no angle dependency is considered in the BLR LOC model here.}
  Nevertheless, the fact that a large portion of the rim below the
  edge has to be shielded argues in favor of the GKR bowl.

  Thus, a bowl-shaped BLR--torus model leads to a consistent view of
the light curve data.

We checked this model further by comparison with previous RM results
of the BLR at much (up to factor 3) lower luminosity (Grier et al.
2012, Pozo et al. 2012, Kollatschny et al. 2014): In all these RM
campaigns between 2008 and 2011 rather small H$\beta$ lags (24--28
days) and a symmetric CCF have been found; Pozo et al. even
found an extremely sharp CCF.  If interpreted in the picture  of a GKR bowl
with a dust rim like that shown in Fig.~\ref{fig:gkr_bowl_250}
(i.e., where the edge lies at $R_x \approx 220$\,ld), then the isodelay
surfaces indicate that the BLR was located at small $R_x \le 50$\,ld.
There the height $H$ of the bowl rim is small
($H / R_x < 15 / 50 \approx 1/3$, see blue long-dashed line in
Fig.~\ref{fig:gkr_bowl_250}).  The same holds for the H$\alpha$ BLR in
2008, with a lag similar to H$\beta$ (Kollatschny et al. 2014).
Finally, a bowl-shaped BLR-dust torus geometry provides a natural way
to obtain -- after correction for the foreshortening effects --
consistent M$_{\rm vir}$ estimates for different Balmer lines and AGN
luminosities.
          
Because it appears hard to achieve such an overall consistent picture
for other geometric configurations, we think that our new data
together with the previous data provide clear evidence favoring a
bowl-shaped BLR and dust geometry for 3C\,120.

\subsection{Consequences for the size/lag -- luminosity relationships}
\label{ssec:discussion_r_l}

Two important size/lag -- luminosity relationships are known: one
between the size of the BLR and the AGN luminosity
(\citealt{2000ApJ...533..631K, 2013ApJ...767..149B}), and one between
the lag of the inner dust torus and the AGN luminosity
(\citealt{1999AstL...25..483O, 2014ApJ...788..159K}).  We note that the
first relation assumes that the BLR size can be estimated by
$c \cdot \tau$.  This assumption holds if the BLR is spherical or
located in the equatorial plane.  However, it does not hold in a
simple manner for a bowl-shaped BLR and torus geometry because then
foreshortening effects play a role.  Therefore, these relations should
actually be considered as lag -- luminosity relations and the
transformation of lag to physical size should be carried out taking the
geometry into account.
   
The relation between dust lag and AGN luminosity shows a scatter by a
factor 2 around the average (solid line in
Fig.~\ref{fig:koshida_comp}).  To understand the scatter, we assume
that all sources in Fig.~\ref{fig:koshida_comp} exhibit a bowl-shaped
torus geometry and that the hot dust emission occurs primarily from
the edge of the bowl-shaped torus (Fig.~\ref{fig:gkr_bowl_250}).  Then
the scatter of the relation may naturally be explained by a hysteresis
effect.  If the AGN quickly brightens but the torus does not respond
quickly enough, then the source moves to the left of the average
relation (as 3C\,120).  On the other hand, if the AGN quickly dims,
we can speculate that the torus dimension shrinks slowly with a
hysteresis, leading to a right-hand shift of the position of the source in
the dust lag -- AGN luminosity diagram.

Furthermore, \cite{2014ApJ...788..159K} compared the reverberation
based inner torus size adopted as $R_{dust} = c \cdot \tau$ with the
size from interferometric measurements.  The interferometric radius in
the $K$ band was found to be systematically larger than the dust
reverberation radius in the same band by about a factor of
two. \cite{2014ApJ...788..159K} proposed that this might be
interpreted by the difference between the flux-weighted radius and the
response-weighted radius of the innermost dust torus.  We point
out that a straightforward natural explanation of this difference is
also provided by the foreshortening effect. If the hot dust emission
occurs primarily from the edge of the bowl-shaped torus
(Fig.~\ref{fig:gkr_bowl_250}), then the lag is about a factor 2--3
shorter than simply inferred from the distance of the edge from the
BH.  The importance of the foreshortening effect was already noted by
\cite{2014A&A...561L...8P} for Sy-1 galaxy WPVS48 (blue source
in Fig.~\ref{fig:koshida_comp}).

Likewise the BLR lag depends on where the BLR is located in the bowl,
for example, further in the bottom or further along the rim at
increased scale height.  As pointed out in
Sect.~\ref{ssec:discussion_bowl_varying}, the bowl size may stay
roughly constant, while the luminosity changes by 30\% but the lag may
suffer from foreshortening.  Then, for a sample of AGN, this
introduces a scatter in the $R-L$ relationship.

So far it is not yet clear whether this effect applies mainly the \Ha
line or whether it also affects H$\beta$ or other broad emission
lines. Certainly, \Ha requires rather a low excitation potential and
therefore the \Ha BLR is more extended than other emission lines
typically used for RM. Therefore, the effects of
the bowl-shaped geometry might be strongest for \Ha.  On the other
hand, \Ha is the brightest broad AGN emission line at optical
wavelengths and as such well suited for PRM.
Future coordinated dust and BLR reverberation campaigns will provide
further insight.

\section{Summary}
\label{ssec:summary}

{We carried out a $BVRIJK$ and 680\,nm NB PRM campaign of \C during
  the epoch 2014 -- 2015. The \Ha line was completely covered by our
  NB, as verified by a contemporaneous spectrum.  By applying the FVG
  method, we separate host and AGN flux in our filters.  We compare
  the BLR and dust reverberation data with a range of bowl-shaped dust
  geometries as first invoked by \cite{2010ApJ...724L.183K}, which
  were subsequently refined by \cite{2012MNRAS.426.3086G} including
  the BLR.  In this view, the BLR clouds are seen only on that side of
  the bowl facing the observer.  The simultaneous BLR and dust
  reverberation data allow us to draw the following conclusions:
    
  \begin{enumerate}
  \item During our epoch, \C has passed a high brightness state with
    $log(\lambda L_{\lambda 5100}) = {44.32}$~erg\,s$^{-1}$, which is three
    times brighter than in 2009/2010.

  \item The CCF between $B$ and \Ha yields a mean
    rest-frame delay of $\tau_{\mathrm{BLR}}=68.9^{+12.4}_{-13.3}\,$ld.
    The CCF is broader than indicated by the rather small
    uncertainties; it shows a broad plateau between 55 and 95 days and
    a clear substructure peak at 60 days.  This indicates that the BLR
    is spread over a large range of isodelay surfaces.

  \item The NIR $J$ and $K$ band light curves allow us
    to measure a dust emission delay
    ($\tau_{K}= 94.4 ^{+4.1}_{-6.9}\,$ld), which is unusually short
    compared to the BLR delay.  Likewise the CCF between $B$ and the
    NIR bands is surprisingly sharper than between $B$ and
    \Ha.  This indicates that the hot dust emission arises from a
    confined volume, which is covered by a narrow range of isodelay
    surfaces.

  \item Our data reach good consistency with the model by
    \cite{2012MNRAS.426.3086G}, where the BLR clouds shield the
    bowl-shaped dust rim up to a half-covering angle of
    $\theta \approx 40^\circ$ from the nuclear radiation. Such a BLR
    reproduces the observed structured broad $B$ -- \Ha CCF with a
    pronounced low-$\tau$ peak predicted from the models.  In
    parallel, the hot dust emission arises from the small part of the
    rim at $40^\circ < \theta < 45^\circ$ facing the observer,
    reproducing the sharp and symmetric $B$ -- NIR CCF; because this
    hot dust edge lies closer to the observer than the equatorial
    plane this also  explains the short lag by foreshortening of the
    light travel time.

  \item In a bowl-shaped geometry the calculation of the viral black
    hole mass M$_{\rm vir}$ may be affected by the foreshortening of
    the observed BLR lag, which leads to an underestimate of the BLR
    distance from the BH and of M$_{\rm vir}$.  The
    foreshortening effect increases with increasing distance of the
    BLR from the BH.  For the 3C\,120 data, a tentative
    correction of the foreshortening effect by spectroscopic or
    geometric $H/R_x$ estimates yields consistent M$_{\rm vir}$ values
    for different Balmer lines (\Ha, \Hb) and different AGN
    luminosities.

  \item If AGN in general exhibit a bowl-shaped BLR-dust torus
    geometry, the foreshortening effects naturally contribute to the
    scatter in the relationship between AGN luminosity and
    lag-inferred BLR size.  The foreshortening effects may also
    explain why reverberation-based torus sizes are on average about a
    factor two smaller compared with interferometric torus sizes.
    Hysteresis effects may also contribute to the scatter in the
    relationship between AGN luminosity and dust lag.
  \end{enumerate}
  Combined with spectral and infrared data, our analysis indicates a
  vertically extended BLR region that can be explained as an envelope
  of a thick dust torus with a bowl-shaped inner border.

  Further light curves and emission line spectra for a sample of
  objects are required to corroborate this approach. The monitoring of
  the BLR emission lines should ideally cover distinct luminosity
  epochs of a single source, which will reveal how the vertical
  structure of a particular emission line grows with luminosity. This
  is especially interesting for the strong and thus easily measurable
  \Ha emission line that likely has the largest variation in scale
  height.}
\begin{acknowledgements}
  We are grateful to Emilio E. Falco and Marco Berton for supporting
  us with recent spectroscopy.  This research has made use of the
  NASA/IPAC Extragalactic Database (NED) which is operated by the Jet
  Propulsion Laboratory, California Institute of Technology, under
  contract with the National Aeronautics and Space Administration.
  This publication is supported as a project of the
  Nordrhein-Westf\"alische Akademie der Wissenschaften und der
  K\"unste in the framework of the academy program by the Federal
  Republic of Germany and the state Nordrhein-Westfalen.  This work is
  supported by the DFG programs Ha\,3555/12-1, Ko\,857/32-2 and
  Ko\,857/33-1.  The observations on Cerro Armazones benefited from
  the care of the guardians Hector Labra, Gerardo Pino, Roberto Munoz,
  and Francisco Arraya.  We thank the referee for critical and
  constructive comments.
\end{acknowledgements}

\bibliographystyle{aa} 
\bibliography{3C120_v11}

\begin{thebibliography}{87}
\expandafter\ifx\csname natexlab\endcsname\relax\def\natexlab#1{#1}\fi

\bibitem[{{Agudo} {et~al.}(2012){Agudo}, {G{\'o}mez}, {Casadio}, {Cawthorne},
  \& {Roca-Sogorb}}]{2012ApJ...752...92A}
{Agudo}, I., {G{\'o}mez}, J.~L., {Casadio}, C., {Cawthorne}, T.~V., \&
  {Roca-Sogorb}, M. 2012, \apj, 752, 92

\bibitem[{{Bahcall} {et~al.}(1972){Bahcall}, {Kozlovsky}, \&
  {Salpeter}}]{1972ApJ...171..467B}
{Bahcall}, J.~N., {Kozlovsky}, B.-Z., \& {Salpeter}, E.~E. 1972, \apj, 171, 467

\bibitem[{{Baldwin} {et~al.}(1995){Baldwin}, {Ferland}, {Korista}, \&
  {Verner}}]{1995ApJ...455L.119B}
{Baldwin}, J., {Ferland}, G., {Korista}, K., \& {Verner}, D. 1995, \apjl, 455,
  L119

\bibitem[{{Barvainis}(1987)}]{1987ApJ...320..537B}
{Barvainis}, R. 1987, \apj, 320, 537

\bibitem[{{Bentz} {et~al.}(2013){Bentz}, {Denney}, {Grier}, {Barth},
  {Peterson}, {Vestergaard}, {Bennert}, {Canalizo}, {De Rosa}, {Filippenko},
  {Gates}, {Greene}, {Li}, {Malkan}, {Pogge}, {Stern}, {Treu}, \&
  {Woo}}]{2013ApJ...767..149B}
{Bentz}, M.~C., {Denney}, K.~D., {Grier}, C.~J., {et~al.} 2013, \apj, 767, 149

\bibitem[{{Bentz} {et~al.}(2009){Bentz}, {Peterson}, {Netzer}, {Pogge}, \&
  {Vestergaard}}]{2009ApJ...697..160B}
{Bentz}, M.~C., {Peterson}, B.~M., {Netzer}, H., {Pogge}, R.~W., \&
  {Vestergaard}, M. 2009, \apj, 697, 160

\bibitem[{{Bentz} {et~al.}(2010){Bentz}, {Walsh}, {Barth}, {Yoshii}, {Woo},
  {Wang}, {Treu}, {Thornton}, {Street}, {Steele}, {Silverman}, {Serduke},
  {Sakata}, {Minezaki}, {Malkan}, {Li}, {Lee}, {Hiner}, {Hidas}, {Greene},
  {Gates}, {Ganeshalingam}, {Filippenko}, {Canalizo}, {Bennert}, \&
  {Baliber}}]{2010ApJ...716..993B}
{Bentz}, M.~C., {Walsh}, J.~L., {Barth}, A.~J., {et~al.} 2010, \apj, 716, 993

\bibitem[{{Bertin}(2006)}]{2006ASPC..351..112B}
{Bertin}, E. 2006, in Astronomical Society of the Pacific Conference Series,
  Vol. 351, Astronomical Data Analysis Software and Systems XV, ed.
  C.~{Gabriel}, C.~{Arviset}, D.~{Ponz}, \& S.~{Enrique}, 112

\bibitem[{{Bertin} \& {Arnouts}(1996)}]{1996A&AS..117..393B}
{Bertin}, E. \& {Arnouts}, S. 1996, \aaps, 117, 393

\bibitem[{{Bertin} {et~al.}(2002){Bertin}, {Mellier}, {Radovich}, {Missonnier},
  {Didelon}, \& {Morin}}]{2002ASPC..281..228B}
{Bertin}, E., {Mellier}, Y., {Radovich}, M., {et~al.} 2002, in Astronomical
  Society of the Pacific Conference Series, Vol. 281, Astronomical Data
  Analysis Software and Systems XI, ed. D.~A. {Bohlender}, D.~{Durand}, \&
  T.~H. {Handley}, 228

\bibitem[{{Berton} {et~al.}(2015){Berton}, {Foschini}, {Ciroi}, {Cracco}, {La
  Mura}, {Lister}, {Mathur}, {Peterson}, {Richards}, \&
  {Rafanelli}}]{2015A&A...578A..28B}
{Berton}, M., {Foschini}, L., {Ciroi}, S., {et~al.} 2015, \aap, 578, A28

\bibitem[{{Choloniewski}(1981)}]{1981AcA....31..293C}
{Choloniewski}, J. 1981, \actaa, 31, 293

\bibitem[{{Collier} {et~al.}(1999){Collier}, {Horne}, {Wanders}, \&
  {Peterson}}]{1999MNRAS.302L..24C}
{Collier}, S., {Horne}, K., {Wanders}, I., \& {Peterson}, B.~M. 1999, \mnras,
  302, L24

\bibitem[{{Collin} \& {Kawaguchi}(2004)}]{2004A&A...426..797C}
{Collin}, S. \& {Kawaguchi}, T. 2004, \aap, 426, 797

\bibitem[{{Collin} {et~al.}(2006){Collin}, {Kawaguchi}, {Peterson}, \&
  {Vestergaard}}]{2006A&A...456...75C}
{Collin}, S., {Kawaguchi}, T., {Peterson}, B.~M., \& {Vestergaard}, M. 2006,
  \aap, 456, 75

\bibitem[{{Czerny} \& {Hryniewicz}(2011)}]{2011A&A...525L...8C}
{Czerny}, B. \& {Hryniewicz}, K. 2011, \aap, 525, L8

\bibitem[{{Czerny} {et~al.}(2013){Czerny}, {Hryniewicz}, {Maity},
  {Schwarzenberg-Czerny}, {{\.Z}ycki}, \& {Bilicki}}]{2013A&A...556A..97C}
{Czerny}, B., {Hryniewicz}, K., {Maity}, I., {et~al.} 2013, \aap, 556, A97

\bibitem[{{Denney} {et~al.}(2009){Denney}, {Peterson}, {Dietrich},
  {Vestergaard}, \& {Bentz}}]{2009ApJ...692..246D}
{Denney}, K.~D., {Peterson}, B.~M., {Dietrich}, M., {Vestergaard}, M., \&
  {Bentz}, M.~C. 2009, \apj, 692, 246

\bibitem[{{Edelson} \& {Krolik}(1988)}]{1988ApJ...333..646E}
{Edelson}, R.~A. \& {Krolik}, J.~H. 1988, \apj, 333, 646

\bibitem[{{Fabricant} {et~al.}(1998){Fabricant}, {Cheimets}, {Caldwell}, \&
  {Geary}}]{1998PASP..110...79F}
{Fabricant}, D., {Cheimets}, P., {Caldwell}, N., \& {Geary}, J. 1998, \pasp,
  110, 79

\bibitem[{{Fioc} \& {Rocca-Volmerange}(1997)}]{1997ASSL..210..257F}
{Fioc}, M. \& {Rocca-Volmerange}, B. 1997, in Astrophysics and Space Science
  Library, Vol. 210, The Impact of Large Scale Near-IR Sky Surveys, ed.
  F.~{Garzon}, N.~{Epchtein}, A.~{Omont}, B.~{Burton}, \& P.~{Persi}, 257

\bibitem[{{Gaskell}(1988)}]{1988ApJ...325..114G}
{Gaskell}, C.~M. 1988, \apj, 325, 114

\bibitem[{{Gaskell} {et~al.}(2007){Gaskell}, {Klimek}, \&
  {Nazarova}}]{2007arXiv0711.1025G}
{Gaskell}, C.~M., {Klimek}, E.~S., \& {Nazarova}, L.~S. 2007, ArXiv e-prints, 
  [eprint {0711.1025}]

\bibitem[{{Goad} {et~al.}(2012){Goad}, {Korista}, \&
  {Ruff}}]{2012MNRAS.426.3086G}
{Goad}, M.~R., {Korista}, K.~T., \& {Ruff}, A.~J. 2012, \mnras, 426, 3086

\bibitem[{{Grier} {et~al.}(2013){Grier}, {Peterson}, {Horne}, {Bentz}, {Pogge},
  {Denney}, {De Rosa}, {Martini}, {Kochanek}, {Zu}, {Shappee}, {Siverd},
  {Beatty}, {Sergeev}, {Kaspi}, {Araya Salvo}, {Bird}, {Bord}, {Borman}, {Che},
  {Chen}, {Cohen}, {Dietrich}, {Doroshenko}, {Efimov}, {Free}, {Ginsburg},
  {Henderson}, {King}, {Mogren}, {Molina}, {Mosquera}, {Nazarov}, {Okhmat},
  {Pejcha}, {Rafter}, {Shields}, {Skowron}, {Szczygiel}, {Valluri}, \& {van
  Saders}}]{2013ApJ...764...47G}
{Grier}, C.~J., {Peterson}, B.~M., {Horne}, K., {et~al.} 2013, \apj, 764, 47

\bibitem[{{Grier} {et~al.}(2012){Grier}, {Peterson}, {Pogge}, {Denney},
  {Bentz}, {Martini}, {Sergeev}, {Kaspi}, {Minezaki}, {Zu}, {Kochanek},
  {Siverd}, {Shappee}, {Stanek}, {Araya Salvo}, {Beatty}, {Bird}, {Bord},
  {Borman}, {Che}, {Chen}, {Cohen}, {Dietrich}, {Doroshenko}, {Drake},
  {Efimov}, {Free}, {Ginsburg}, {Henderson}, {King}, {Koshida}, {Mogren},
  {Molina}, {Mosquera}, {Nazarov}, {Okhmat}, {Pejcha}, {Rafter}, {Shields},
  {Skowron}, {Szczygiel}, {Valluri}, \& {van Saders}}]{2012ApJ...755...60G}
{Grier}, C.~J., {Peterson}, B.~M., {Pogge}, R.~W., {et~al.} 2012, \apj, 755, 60

\bibitem[{{Haas} {et~al.}(2011){Haas}, {Chini}, {Ramolla}, {Pozo Nu{\~n}ez},
  {Westhues}, {Watermann}, {Hoffmeister}, \& {Murphy}}]{2011A&A...535A..73H}
{Haas}, M., {Chini}, R., {Ramolla}, M., {et~al.} 2011, \aap, 535, A73

\bibitem[{{Haas} {et~al.}(2012){Haas}, {Hackstein}, {Ramolla}, {Drass},
  {Watermann}, {Lemke}, \& {Chini}}]{2012AN....333..706H}
{Haas}, M., {Hackstein}, M., {Ramolla}, M., {et~al.} 2012, Astronomische
  Nachrichten, 333, 706

\bibitem[{{Ho} {et~al.}(1997){Ho}, {Filippenko}, {Sargent}, \&
  {Peng}}]{1997ApJS..112..391H}
{Ho}, L.~C., {Filippenko}, A.~V., {Sargent}, W.~L.~W., \& {Peng}, C.~Y. 1997,
  \apjs, 112, 391

\bibitem[{{Hodapp} {et~al.}(2010){Hodapp}, {Chini}, {Reipurth}, {Murphy},
  {Lemke}, {Watermann}, {Jacobson}, {Bischoff}, {Chonis}, {Dement}, {Terrien},
  {Bott}, \& {Provence}}]{2010SPIE.7735E..1AH}
{Hodapp}, K.~W., {Chini}, R., {Reipurth}, B., {et~al.} 2010, in Society of
  Photo-Optical Instrumentation Engineers (SPIE) Conference Series, Vol. 7735,
  Society of Photo-Optical Instrumentation Engineers (SPIE) Conference Series,
  1

\bibitem[{{H{\"o}nig} {et~al.}(2017){H{\"o}nig}, {Watson}, {Kishimoto},
  {Gandhi}, {Goad}, {Horne}, {Shankar}, {Banerji}, {Boulderstone}, {Jarvis},
  {Smith}, \& {Sullivan}}]{2017MNRAS.464.1693H}
{H{\"o}nig}, S.~F., {Watson}, D., {Kishimoto}, M., {et~al.} 2017, \mnras, 464,
  1693

\bibitem[{{Horne} {et~al.}(2004){Horne}, {Peterson}, {Collier}, \&
  {Netzer}}]{2004PASP..116..465H}
{Horne}, K., {Peterson}, B.~M., {Collier}, S.~J., \& {Netzer}, H. 2004, \pasp,
  116, 465

\bibitem[{{Horne} {et~al.}(1991){Horne}, {Welsh}, \&
  {Peterson}}]{1991ApJ...367L...5H}
{Horne}, K., {Welsh}, W.~F., \& {Peterson}, B.~M. 1991, \apjl, 367, L5

\bibitem[{{Huchra} \& {Burg}(1992)}]{1992ApJ...393...90H}
{Huchra}, J. \& {Burg}, R. 1992, \apj, 393, 90

\bibitem[{{Kabath} {et~al.}(2009){Kabath}, {Erikson}, {Rauer}, {Pasternacki},
  {Csizmadia}, {Chini}, {Lemke}, {Murphy}, {Fruth}, {Titz}, \&
  {Eigm{\"u}ller}}]{2009A&A...506..569K}
{Kabath}, P., {Erikson}, A., {Rauer}, H., {et~al.} 2009, \aap, 506, 569

\bibitem[{{Kaspi} {et~al.}(2000){Kaspi}, {Smith}, {Netzer}, {Maoz}, {Jannuzi},
  \& {Giveon}}]{2000ApJ...533..631K}
{Kaspi}, S., {Smith}, P.~S., {Netzer}, H., {et~al.} 2000, \apj, 533, 631

\bibitem[{{Kawaguchi} \& {Mori}(2010)}]{2010ApJ...724L.183K}
{Kawaguchi}, T. \& {Mori}, M. 2010, \apjl, 724, L183

\bibitem[{{Kawaguchi} \& {Mori}(2011)}]{2011ApJ...737..105K}
{Kawaguchi}, T. \& {Mori}, M. 2011, \apj, 737, 105

\bibitem[{{Kishimoto} {et~al.}(2013){Kishimoto}, {H{\"o}nig}, {Antonucci},
  {Millan-Gabet}, {Barvainis}, {Millour}, {Kotani}, {Tristram}, \&
  {Weigelt}}]{2013ApJ...775L..36K}
{Kishimoto}, M., {H{\"o}nig}, S.~F., {Antonucci}, R., {et~al.} 2013, \apjl,
  775, L36

\bibitem[{{Kishimoto} {et~al.}(2007){Kishimoto}, {H{\"o}nig}, {Beckert}, \&
  {Weigelt}}]{2007A&A...476..713K}
{Kishimoto}, M., {H{\"o}nig}, S.~F., {Beckert}, T., \& {Weigelt}, G. 2007,
  \aap, 476, 713

\bibitem[{{Kollatschny}(2003)}]{2003A&A...412L..61K}
{Kollatschny}, W. 2003, \aap, 412, L61

\bibitem[{{Kollatschny} {et~al.}(2014){Kollatschny}, {Ulbrich}, {Zetzl},
  {Kaspi}, \& {Haas}}]{2014A&A...566A.106K}
{Kollatschny}, W., {Ulbrich}, K., {Zetzl}, M., {Kaspi}, S., \& {Haas}, M. 2014,
  \aap, 566, A106

\bibitem[{{Kollatschny} \& {Zetzl}(2011)}]{2011Natur.470..366K}
{Kollatschny}, W. \& {Zetzl}, M. 2011, \nat, 470, 366

\bibitem[{{Kollatschny} \& {Zetzl}(2013{\natexlab{a}})}]{2013A&A...549A.100K}
{Kollatschny}, W. \& {Zetzl}, M. 2013{\natexlab{a}}, \aap, 549, A100

\bibitem[{{Kollatschny} \& {Zetzl}(2013{\natexlab{b}})}]{2013A&A...558A..26K}
{Kollatschny}, W. \& {Zetzl}, M. 2013{\natexlab{b}}, \aap, 558, A26

\bibitem[{{Koshida} {et~al.}(2014){Koshida}, {Minezaki}, {Yoshii}, {Kobayashi},
  {Sakata}, {Sugawara}, {Enya}, {Suganuma}, {Tomita}, {Aoki}, \&
  {Peterson}}]{2014ApJ...788..159K}
{Koshida}, S., {Minezaki}, T., {Yoshii}, Y., {et~al.} 2014, \apj, 788, 159

\bibitem[{{Landolt}(2009)}]{2009AJ....137.4186L}
{Landolt}, A.~U. 2009, \aj, 137, 4186

\bibitem[{{Landt} {et~al.}(2008){Landt}, {Bentz}, {Ward}, {Elvis}, {Peterson},
  {Korista}, \& {Karovska}}]{2008ApJS..174..282L}
{Landt}, H., {Bentz}, M.~C., {Ward}, M.~J., {et~al.} 2008, \apjs, 174, 282

\bibitem[{{Landt} {et~al.}(2011){Landt}, {Elvis}, {Ward}, {Bentz}, {Korista},
  \& {Karovska}}]{2011MNRAS.414..218L}
{Landt}, H., {Elvis}, M., {Ward}, M.~J., {et~al.} 2011, \mnras, 414, 218

\bibitem[{{Laor}(2004)}]{2004ASPC..311..169L}
{Laor}, A. 2004, in Astronomical Society of the Pacific Conference Series, Vol.
  311, AGN Physics with the Sloan Digital Sky Survey, ed. G.~T. {Richards} \&
  P.~B. {Hall}, 169

\bibitem[{{Lawrence}(1991)}]{1991MNRAS.252..586L}
{Lawrence}, A. 1991, \mnras, 252, 586

\bibitem[{{Maoz} {et~al.}(1991){Maoz}, {Netzer}, {Mazeh}, {Beck}, {Almoznino},
  {Leibowitz}, {Brosch}, {Mendelson}, \& {Laor}}]{1991ApJ...367..493M}
{Maoz}, D., {Netzer}, H., {Mazeh}, T., {et~al.} 1991, \apj, 367, 493

\bibitem[{{Michel} \& {Huchra}(1988)}]{1988PASP..100.1423M}
{Michel}, A. \& {Huchra}, J. 1988, \pasp, 100, 1423

\bibitem[{{Mor} \& {Trakhtenbrot}(2011)}]{2011ApJ...737L..36M}
{Mor}, R. \& {Trakhtenbrot}, B. 2011, \apjl, 737, L36

\bibitem[{{Oknyanskij}(1999)}]{1999OAP....12...99O}
{Oknyanskij}, V.~L. 1999, Odessa Astronomical Publications, 12, 99

\bibitem[{{Oknyanskij} {et~al.}(1999){Oknyanskij}, {Lyuty}, {Taranova}, \&
  {Shenavrin}}]{1999AstL...25..483O}
{Oknyanskij}, V.~L., {Lyuty}, V.~M., {Taranova}, O.~G., \& {Shenavrin}, V.~I.
  1999, Astronomy Letters, 25, 483

\bibitem[{{Onken} {et~al.}(2004){Onken}, {Ferrarese}, {Merritt}, {Peterson},
  {Pogge}, {Vestergaard}, \& {Wandel}}]{2004ApJ...615..645O}
{Onken}, C.~A., {Ferrarese}, L., {Merritt}, D., {et~al.} 2004, \apj, 615, 645

\bibitem[{{Osterbrock}(1978)}]{1978PNAS...75..540O}
{Osterbrock}, D.~E. 1978, Proceedings of the National Academy of Science, 75,
  540

\bibitem[{{Pancoast} {et~al.}(2012){Pancoast}, {Brewer}, {Treu}, {Barth},
  {Bennert}, {Canalizo}, {Filippenko}, {Gates}, {Greene}, {Li}, {Malkan},
  {Sand}, {Stern}, {Woo}, {Assef}, {Bae}, {Buehler}, {Cenko}, {Clubb},
  {Cooper}, {Diamond-Stanic}, {Hiner}, {H{\"o}nig}, {Joner}, {Kandrashoff},
  {Laney}, {Lazarova}, {Nierenberg}, {Park}, {Silverman}, {Son}, {Sonnenfeld},
  {Thorman}, {Tollerud}, {Walsh}, \& {Walters}}]{2012ApJ...754...49P}
{Pancoast}, A., {Brewer}, B.~J., {Treu}, T., {et~al.} 2012, \apj, 754, 49

\bibitem[{{Patat} {et~al.}(2011){Patat}, {Moehler}, {O'Brien}, {Pompei},
  {Bensby}, {Carraro}, {de Ugarte Postigo}, {Fox}, {Gavignaud}, {James},
  {Korhonen}, {Ledoux}, {Randall}, {Sana}, {Smoker}, {Stefl}, \&
  {Szeifert}}]{2011A&A...527A..91P}
{Patat}, F., {Moehler}, S., {O'Brien}, K., {et~al.} 2011, \aap, 527, A91

\bibitem[{{Peterson}(1993)}]{1993PASP..105..247P}
{Peterson}, B.~M. 1993, \pasp, 105, 247

\bibitem[{{Peterson} {et~al.}(2004){Peterson}, {Ferrarese}, {Gilbert}, {Kaspi},
  {Malkan}, {Maoz}, {Merritt}, {Netzer}, {Onken}, {Pogge}, {Vestergaard}, \&
  {Wandel}}]{2004ApJ...613..682P}
{Peterson}, B.~M., {Ferrarese}, L., {Gilbert}, K.~M., {et~al.} 2004, \apj, 613,
  682

\bibitem[{{Peterson} \& {Wandel}(1999)}]{1999ApJ...521L..95P}
{Peterson}, B.~M. \& {Wandel}, A. 1999, \apjl, 521, L95

\bibitem[{{Peterson} {et~al.}(1998){Peterson}, {Wanders}, {Horne}, {Collier},
  {Alexander}, {Kaspi}, \& {Maoz}}]{1998PASP..110..660P}
{Peterson}, B.~M., {Wanders}, I., {Horne}, K., {et~al.} 1998, \pasp, 110, 660

\bibitem[{{Pozo Nu{\~n}ez} {et~al.}(2014{\natexlab{a}}){Pozo Nu{\~n}ez},
  {Haas}, {Chini}, {Ramolla}, {Westhues}, {Steenbrugge}, {Kaderhandt}, {Drass},
  {Lemke}, \& {Murphy}}]{2014A&A...561L...8P}
{Pozo Nu{\~n}ez}, F., {Haas}, M., {Chini}, R., {et~al.} 2014{\natexlab{a}},
  \aap, 561, L8

\bibitem[{{Pozo Nu{\~n}ez} {et~al.}(2014{\natexlab{b}}){Pozo Nu{\~n}ez},
  {Haas}, {Ramolla}, {Bruckmann}, {Westhues}, {Chini}, {Steenbrugge}, {Lemke},
  {Murphy}, \& {Kollatschny}}]{2014A&A...568A..36P}
{Pozo Nu{\~n}ez}, F., {Haas}, M., {Ramolla}, M., {et~al.} 2014{\natexlab{b}},
  \aap, 568, A36

\bibitem[{{Pozo Nu{\~n}ez} {et~al.}(2012){Pozo Nu{\~n}ez}, {Ramolla},
  {Westhues}, {Bruckmann}, {Haas}, {Chini}, {Steenbrugge}, \&
  {Murphy}}]{2012A&A...545A..84P}
{Pozo Nu{\~n}ez}, F., {Ramolla}, M., {Westhues}, C., {et~al.} 2012, \aap, 545,
  A84

\bibitem[{{Pozo Nu{\~n}ez} {et~al.}(2015){Pozo Nu{\~n}ez}, {Ramolla},
  {Westhues}, {Haas}, {Chini}, {Steenbrugge}, {Barr Dom{\'{\i}}nguez},
  {Kaderhandt}, {Hackstein}, {Kollatschny}, {Zetzl}, {Hodapp}, \&
  {Murphy}}]{2015A&A...576A..73P}
{Pozo Nu{\~n}ez}, F., {Ramolla}, M., {Westhues}, C., {et~al.} 2015, \aap, 576,
  A73

\bibitem[{{Pozo Nu{\~n}ez} {et~al.}(2013){Pozo Nu{\~n}ez}, {Westhues},
  {Ramolla}, {Bruckmann}, {Haas}, {Chini}, {Steenbrugge}, {Lemke}, \&
  {Murphy}}]{2013A&A...552A...1P}
{Pozo Nu{\~n}ez}, F., {Westhues}, C., {Ramolla}, M., {et~al.} 2013, \aap, 552,
  A1

\bibitem[{{Pringle}(1981)}]{1981ARA&A..19..137P}
{Pringle}, J.~E. 1981, \araa, 19, 137

\bibitem[{{Ramolla} {et~al.}(2013){Ramolla}, {Drass}, {Lemke}, {Westhues},
  {Pozo Nu{\~n}ez}, {Barr Dominguez}, {Haas}, {Chini}, \&
  {Murphy}}]{2013AN....334.1115R}
{Ramolla}, M., {Drass}, H., {Lemke}, R., {et~al.} 2013, Astronomische
  Nachrichten, 334, 1115

\bibitem[{{Ramolla} {et~al.}(2015){Ramolla}, {Pozo Nu{\~n}ez}, {Westhues},
  {Haas}, \& {Chini}}]{2015A&A...581A..93R}
{Ramolla}, M., {Pozo Nu{\~n}ez}, F., {Westhues}, C., {Haas}, M., \& {Chini}, R.
  2015, \aap, 581, A93

\bibitem[{{Sakata} {et~al.}(2010){Sakata}, {Minezaki}, {Yoshii}, {Kobayashi},
  {Koshida}, {Aoki}, {Enya}, {Tomita}, {Suganuma}, {Katsuno Uchimoto}, \&
  {Sugawara}}]{2010ApJ...711..461S}
{Sakata}, Y., {Minezaki}, T., {Yoshii}, Y., {et~al.} 2010, \apj, 711, 461

\bibitem[{{Schlafly} \& {Finkbeiner}(2011)}]{2011ApJ...737..103S}
{Schlafly}, E.~F. \& {Finkbeiner}, D.~P. 2011, \apj, 737, 103

\bibitem[{{Sergeev} {et~al.}(2005){Sergeev}, {Doroshenko}, {Golubinskiy},
  {Merkulova}, \& {Sergeeva}}]{2005ApJ...622..129S}
{Sergeev}, S.~G., {Doroshenko}, V.~T., {Golubinskiy}, Y.~V., {Merkulova},
  N.~I., \& {Sergeeva}, E.~A. 2005, \apj, 622, 129

\bibitem[{{Shakura} \& {Sunyaev}(1973)}]{1973A&A....24..337S}
{Shakura}, N.~I. \& {Sunyaev}, R.~A. 1973, \aap, 24, 337

\bibitem[{{Storey} \& {Zeippen}(2000)}]{2000MNRAS.312..813S}
{Storey}, P.~J. \& {Zeippen}, C.~J. 2000, \mnras, 312, 813

\bibitem[{{Suganuma} {et~al.}(2006){Suganuma}, {Yoshii}, {Kobayashi},
  {Minezaki}, {Enya}, {Tomita}, {Aoki}, {Koshida}, \&
  {Peterson}}]{2006ApJ...639...46S}
{Suganuma}, M., {Yoshii}, Y., {Kobayashi}, Y., {et~al.} 2006, \apj, 639, 46

\bibitem[{{Sulentic} {et~al.}(2000){Sulentic}, {Zwitter}, {Marziani}, \&
  {Dultzin-Hacyan}}]{2000ApJ...536L...5S}
{Sulentic}, J.~W., {Zwitter}, T., {Marziani}, P., \& {Dultzin-Hacyan}, D. 2000,
  \apjl, 536, L5

\bibitem[{{Tokarz} \& {Roll}(1997)}]{1997ASPC..125..140T}
{Tokarz}, S.~P. \& {Roll}, J. 1997, in Astronomical Society of the Pacific
  Conference Series, Vol. 125, Astronomical Data Analysis Software and Systems
  VI, ed. G.~{Hunt} \& H.~{Payne}, 140

\bibitem[{{Urry} \& {Padovani}(1995)}]{1995PASP..107..803U}
{Urry}, C.~M. \& {Padovani}, P. 1995, \pasp, 107, 803

\bibitem[{{Vestergaard}(2002)}]{2002ApJ...571..733V}
{Vestergaard}, M. 2002, \apj, 571, 733

\bibitem[{{Watson} {et~al.}(2011){Watson}, {Denney}, {Vestergaard}, \&
  {Davis}}]{2011ApJ...740L..49W}
{Watson}, D., {Denney}, K.~D., {Vestergaard}, M., \& {Davis}, T.~M. 2011,
  \apjl, 740, L49

\bibitem[{{Welsh}(1999)}]{1999PASP..111.1347W}
{Welsh}, W.~F. 1999, \pasp, 111, 1347

\bibitem[{{Winkler} {et~al.}(1992){Winkler}, {Glass}, {van Wyk}, {Marang},
  {Jones}, {Buckley}, \& {Sekiguchi}}]{1992MNRAS.257..659W}
{Winkler}, H., {Glass}, I.~S., {van Wyk}, F., {et~al.} 1992, \mnras, 257, 659

\bibitem[{{Woo} {et~al.}(2007){Woo}, {Treu}, {Malkan}, {Ferry}, \&
  {Misch}}]{2007ApJ...661...60W}
{Woo}, J.-H., {Treu}, T., {Malkan}, M.~A., {Ferry}, M.~A., \& {Misch}, T. 2007,
  \apj, 661, 60

\bibitem[{{Yoshii} {et~al.}(2014){Yoshii}, {Kobayashi}, {Minezaki}, {Koshida},
  \& {Peterson}}]{2014ApJ...784L..11Y}
{Yoshii}, Y., {Kobayashi}, Y., {Minezaki}, T., {Koshida}, S., \& {Peterson},
  B.~A. 2014, \apjl, 784, L11

\end{thebibliography}

\end{document}